\documentclass[12pt]{iopart}

\usepackage{mathtools}
\usepackage{amssymb,mathrsfs,bm,mathdots,xspace,bbold,version}   
\usepackage{amsmath}
\usepackage{graphicx}

\usepackage{tensor}

\usepackage{tikz}

\def\BB{{\cal B}}
\def\CC{{\cal C}}

\def\HH{{\cal H}}

\def\NN{{\cal N}}

\def\Nbb{{\mathbb N}}
\newcommand{\nbb}[1]{\Nbb_{\bar #1}}
\newcommand{\nbbs}[1]{\Nbb^*_{\bar #1}}

\def\ww{{\mathrm{w}}}

\def\Rbb{{\mathbb R}}

\def\Cbb{{\mathbb C}}

\def\Sbb{{\mathbb S}}

\def\Abf{{\mathbf A}}
\def\Tbf{{\mathbf T}}
\def\Fbf{{\mathbf F}}
\def\Bbf{{\mathbf B}}
\def\Ubf{{\mathbf U}}
\def\Dbf{{\mathbf D}}

\def\Bee{\begin{equation}}
\def\Eee{\end{equation}}

\def\Beq{\begin{eqnarray}}
\def\Eeq{\end{eqnarray}}

\def\moyenne#1{\langle#1\rangle}

\def\id{\mathbb{1}}

\usepackage{soul}

\newcommand{\ket}[1]{|#1\rangle}
\newcommand{\bra}[1]{\langle #1|}

\begin{document}

\title[Quantum Chaos Game]{Quantum Repeated Interactions and the Chaos Game}

\author{T. Platini and R. J. Low}

\address{Applied Mathematics Research Center, Coventry University, Coventry, CV1 5FB, England,}

\ead{thierry.platini@coventry.ac.uk}
\vspace{10pt}
\begin{indented}
\item[]\date{\today}
\end{indented}

\begin{abstract}
Inspired by the algorithm of Barnsley's chaos game, we construct an open quantum system model based on the repeated interaction process. We shown that the quantum dynamics of the appropriate fermionic/bosonic system (in interaction with an environment) provides a physical model of the chaos game. When considering fermionic operators, we follow the system's evolution by focusing on its reduced density matrix. The system is shown to be in a Gaussian state (at all time $t$) and the average number of particles is shown to obey the chaos game equation. Considering bosonic operators, with a system initially prepared in coherent states, the evolution of the system can be tracked by investigating the dynamics of the eigenvalues of the annihilation operator. This quantity is governed by a chaos game-like equation from which different scenarios emerge.

\end{abstract}




%
%
%
%
%

\section{Introduction}
Soon after their popularization by Mandelbrot \cite{Mandelbrot_77,Mandelbrot_82}, fractals were the subject of intensive studies for their intrinsic mathematical interest \cite{Barnsley_88,Barnsley_06,Falconer}, for their scientific applications \cite{Birdi,Hastings}, and for the sheer joy of their aesthetic properties \cite{Peitgen}. 

The relationship between fractals and dynamical systems is profound, and a particularly beautiful illustration of this is found in Barnsley's chaos game, in which a simple random dynamical system known as an iterated function system (IFS) and based on three vertices produces the Sierpinski triangle \cite{Barnsley_88}. {In statistical mechanics, the Sierpinski triangle appeared in the study of self-organising systems and cellular automata \cite{Wolfram_83}}. The chaos game itself has found application in the field of genetics starting with \cite{Jeffrey_90} and continuing to the time of writing in, for example \cite{Yin}.  Fractals have also appeared in quantum mechanics in various ways, ranging from their appearance in the path integal formulation of quantum mechanics, implicit in \cite{Feynman} (though fractal terminology had not yet been developed), through the investigation of fractons \cite{Alexander} and even attempts to model quantum space-time \cite{Loll}.

In this work, we will quickly review the rules defining the chaos game, focusing on two well known examples: the $1$d case (with two vertices) leading to the Smith-Volterra-Cantor set and the $2$d case (built around three vertices) leading to the Sierpinski triangle. With the chaos game algorithm in mind, we propose the construction of a model for open quantum system based on the repeated interaction process (section \ref{QCG}). In this model, the system and the bath are described by either fermionic or bosonic quadratic Hamiltonians 
(see sections \ref{The system} and \ref{The environment}). One can easily track the time evolution of the system's operators and follow the system's journey under the bath's influence. Studies for fermionic and bosonic cases are presented in sections \ref{fermion_section} and \ref{boson_section}.

\section{The ``Classical'' Chaos Game}\label{classical}
The chaos game is a mathematical game proposed by Michael Fielding Barnsley in 1988. It is defined by a set of rules, the (non-deterministic)  algorithm which when repeated with appropriate parameter values leads to fractal structures. The game provides an introduction to iterated function systems (IFS) and their attractors. Here we only give a short presentation of the game. A full description of the problem and the mathematics behind it can be found in \cite{Barnsley_88, Barnsley_06}. We start with a set of preparation steps (1) before giving the actual playing rules (2): the algorithm.
\begin{enumerate}
\item[1] Prepare for the game
\begin{enumerate}
\item[1.1] Picture any $M$-gon in $\Rbb^2$ and label its vertices $b_j$ ($j\in\{0,1,...,M-1\}$). 
\item[1.2] Choose another point $x_0$ in $\Rbb^2$. This is the starting point of the game.
\item[1.3] Choose $\ww$ a real number in $]0,1[$.
\end{enumerate}

\item[2] Start to play
\begin{enumerate}
\item[2.1] Select at random one of the vertices of the $M$-gon, call it $c_0$.
\item[2.2] Draw a point $x_{1}$ at position $(1-\ww)x_0+\ww c_0$.
\item[2.3] Repeat steps $[2.1]$ and $[2.2]$ defining $c_n$ and $x_n$ by iteration.
\end{enumerate}
\end{enumerate}
{
We will refer to the chaos game equation as
\Bee\label{chaos_equation}
x_{n+1}=(1-\ww)x_n+\ww c_n,
\Eee
where $c_n$ are independent random variables taking values in $\{b_j\}_{j=0}^{M-1}$, with equal probability.} At this point, it is convenient to define the functions $f_{j}$ from $\Rbb^2\rightarrow\Rbb^2$ by 
\Bee
f_j(x)=(1-\ww)x+\ww b_j \ {\rm for} \ j=0,1,..., M-1.
\Eee
Those are contractions on $\Rbb^2$ so that there exists a unique non-empty compact set $\CC$ that is invariant under $f_j$ \cite{Falconer}. In other words $\CC$ is the set satisfying $\CC=\bigcup_{j}f_j(\CC)$. The set $\CC$ acts as an attractor: for almost all $x_0$, the orbit $\{x_n\}_{n\in\Nbb}$ approaches $\CC$. Defining, for any non-empty compact subset $\Sbb$, $f(\Sbb)=\bigcup_{j}f_j(\Sbb)$ we write $f^k(\Sbb)=f(f^{k-1}(\Sbb))$ with $f^0(\Sbb)=\Sbb$. It follows that $\CC=\bigcap_{k=1}^\infty f^k(\Sbb)$. In fact, the sequence $f^k(\Sbb)$ converges to $\CC$ for any $\Sbb$. We call $f^k(\Sbb)$ a pre-fractal of $\CC$.

\subsection{The $1$d case and Smith-Volterra-Cantor set}
Let us here restrict the chaos game to $\Rbb$ instead of $\Rbb^2$. We replace the $M$-gon by a $2$-gon with $b_0=0$ and $b_1=1$. The functions $f_0$ and $f_1$ are defined from $\Rbb\rightarrow\Rbb$. As an example let us consider $0\le x<y\le 1$ and write $f([x,y])=\cup_j[f_j(x),f_j(y)]$. For $1/2<\ww <1$, it is clear that $f([x,y])$ is the union of two non-intersecting sets. It follows that $f^k([x,y])$ is the union of $2^k$ disjoint subsets. The series $\{f^k(\Sbb)\}_{k\in\Nbb}$ converges towards a well known fractal set called the Smith-Volterra-Cantor set. For $0<\ww <1/2$ one see that $f([0,1])=[0,1]$ so that $[0,1]$ is the unique invariant under $f$. No fractal is emerging from this range of parameter. However, we will investigate this situation in more details when presenting results for the quantum game.

\subsection{The $2$d case and Sierpinski triangle}
Back in $\Rbb^2$, the most famous chaos game example is obtained by considering a triangle ($M=3$ and $b_j=(\cos \theta_j,\sin \theta_j)$ with $\theta_j=2\pi j/3$ for $j=0,1,2$) together with $\ww =1/2$. One observes that, for this system the set $\CC$ is the Sierpinksi triangle. After a large number of iterations, independently of the starting point $x_0$, the set $\{x_n\}_{n\in\Nbb}$ is indeed attracted towards it.
As in the $1d$ case for small values of $\ww $ (by which we mean $0<\ww <1/3$) the density of point appears to be continuous in the subset of $\Rbb^2$ delimited by the triangle $(b_0,b_1,b_2)$. 


\section{The Quantum Chaos Game}\label{QCG}
In this section, we present the set-up (for fermionic and bosonic systems) leading to what we have  called ``the quantum chaos game". 
In both, the fermionic and bosonic scenarios, the set-up is based on a randomised version of the repeated interaction process \cite{Attal_06,Attal_07}. The latter process allows for the description of a quantum system open to interactions with an environment. The ensemble system plus environment is described as a closed quantum system. This approach has been the subject of an active line of research with various applications including quantum trajectories \cite{Bauer_11,Bauer_13,Pellegrini_08,Pellegrini_10}, relaxation, thermalization and transport in quantum systems \cite{Dhar_08,Bruneau_09,Karevski_09,Platini_10}. We will see how the starting point $x_0$ (in the classical chaos game) is associated (in the quantum chaos game) to the initial state of the system. Identically, all vertices $b_j$ of the $M$-gon are associated to the different initial states populating the environment. For both the fermionic and bosonic cases, the Hamiltonians describing the  system-environment interaction are chosen to be quadratic. Apart from being different by the nature of the operators considered, the fermionic and bosonic scenarios differ in the choices of initial states in which different subsystems are prepared. Fermionic subsystems are prepared in one of the canonical states $\ket{1}$, $\ket{0}$ with one or no particles, while bosonic subsystems are prepared in coherent states (eigenstates of the annihilation operator). We will show that for the fermionic case the quantum game is one-dimensional only. It perfectly maps onto the $2$-vertex problem presented in the previous section. In the bosonic case we show how we recover the $2d$ chaos game for arbitrarily shaped $M$-gon. The following subsections presents the general set-up as well as the repeated interaction process for both particle types. We start with a detailed description of the system, the environment, and the initial state (see sections \ref{The system}, \ref{The environment} and \ref{The initial state}). We pursue by describing the interaction between the system and the environment as well as the dynamics governing the evolution of the ensemble (see sections \ref{Hamiltonian} and \ref{The dynamics}).
\subsection{The system} 
\label{The system}
Let us start by defining the system, with Hamiltonian: 
\Bee
H_0=\omega a_0^\dag a_0, 
\Eee
with creation and annihilation operators ($a_0^\dag$ and  $a_0$) which are either bosonic or fermionic operators.
We denote by $\HH_0$ the Hilbert space of the system and $\ket{\phi}_0$ a state of this Hilbert space. When working explicitly with fermions or bosons we will use the notations $f_0$, $f_0^\dag$ or $b_0$, $b_0^\dag$ respectively.
\subsection{The environment}
\label{The environment} We continue by defining the environment as  a bath made of an infinite collection of independent modes $j$ ($j\in\Nbb^*$), with Hilbert space $\HH_j$. We denote by $\ket{\phi}_j$ a state of $\HH_j$ and write $a_j$ and $a^\dag_j$ the associated annihilation and creation operators. Since the modes are independent  the Hamiltonian of the environment is $H_\BB=\sum_{j\in\Nbb^*}H_j$, with $[H_j,H_{i}]=0$ and
\Bee
H_j=\omega a^\dag_j a_j, \ \forall j\in\Nbb^*.
\Eee 
The total Hilbert space associated to the environment is $\BB=\bigotimes_{j\in\Nbb^*}\HH_j$. Recall that the notation $f_j$, $f_j^\dag$ or $b_j$, $b_j^\dag$ will be adopted when working explicitly with fermions or bosons. \footnote{Note that we have chosen the quanta of energy $\omega$ in the environment and the system to be identical. It is of course possible to consider an inhomogeneous set-up by defining $\omega_j$ for all $j\in\Nbb$.}
\subsection{The initial state} 
\label{The initial state} The system is prepared in a state $\ket{\chi_0}_0$, while all modes of the bath ($j\in\Nbb^*$) are prepared independently. To be more explicit we define $M$ particular states of some mode as $\ket{\beta_k}$ for $k\in\{0,1,2,...,M-1\}$ (in general we do not require those states to be orthogonal). In addition, we define the random variables $\gamma_j$ (for $j\in\Nbb^*$), each $\gamma_j$ taking value in $\{0,1,2,...,M-1\}$. We denote by $\ket{\gamma_j}_j$ the initial state of the $j^{\rm{th}}$ mode. It is defined to be one of the $\ket{\beta_k}$ states selected by the random variable $\gamma_j$, so that
\Bee
\ket{\gamma_j}_j=\sum_{k=0}^{M-1}\delta_{\gamma_j,k}\ket{\beta_k}_j,
\Eee
with $\delta_{q,k}=1$ if $k=q$ and zero otherwise. Finally the initial state of the environment is simply $\ket{\Gamma}_\BB=\bigotimes_{j\in\Nbb^*}\ket{\gamma_j}_j$ and overall, the full initial state $\ket{I}$ of the ensemble system-environment is $\ket{I}=\ket{\chi_0}_0\otimes\ket{\Gamma}_\BB$.
\subsection{Hamiltonian} 
\label{Hamiltonian} The dynamics is specified by the full Hamiltonian describing the interaction between the system and the environment using the repeated interaction process \cite{Attal_06}. The Hamiltonian is time dependent and written $H(t)$. We denote by $\tau$ the characteristic time of interaction. Over the time interval $[(n-1)\tau,n\tau[$ the Hamiltonian $H(t)$ is constant and defined by $H(t)= H_{(n)}$ with
\Bee
H_{(n)}
=H_{0}+V_{0,n}+H_\BB, \ \forall t \in [(n-1)\tau,n\tau[,
\Eee
and 
\Bee
V_{0,n}=-\lambda(a_0^\dag a_n+a^\dag_n a_0).
\Eee
Let us write $\nbbs{n}=\Nbb^*\backslash \{n\}$. It is often convenient to write $H_{(n)}=H_{0,n}+\sum_{j\in\nbbs{n}}H_j$ with $H_{0,n}=H_0+V_{0,n}+H_n$ where $[H_{(n)},\sum_{j\in\nbbs{n}}H_j]=0$. To introduce the more compact notation $H_{0,n}=\Abf_n^\dag \Tbf \Abf_n$, we define
\begin{equation}
\label{matT}
\Tbf
=
\left(
\begin{array}{cc}
\omega&-\lambda\\
-\lambda&\omega
\end{array}
\right),
\end{equation}
$\Abf_n^\dag=(a_0^\dag,a_n^\dag)$ and $\Abf_n=(a_0,a_n)^T$.
We will use the notations $\Fbf_n=(f_0,f_n)^T$ and $\Bbf_n=(b_0,b_n)^T$ for fermions and bosons respectively. \footnote{If the reader chooses to consider inhomogeneous $\omega$ values (as indicated before) a matrix $\Tbf_n$ needs to be defined:
$
\Tbf_n
=
\left(
\begin{array}{cc}
\omega_0&-\lambda\\
-\lambda&\omega_n
\end{array}
\right).
$
}
\subsection{The dynamics} 
\label{The dynamics} We now define the time evolution operator $U_{(n)}(\tau)=\exp(-i\tau H_{(n)})$ so that the evolution of the state for the ensemble system-environment is given by
\Bee
\ket{(n+1)\tau}=U_{(n+1)}(\tau)\ket{n\tau},\ \text{with}\ \ket{\tau}=U_{(1)}(\tau)\ket{I},
\Eee
where $\ket{I}$ is the intial state defined in section \ref{The initial state}. Note that 
\Bee
U_{(n)}=\exp(-i\tau H_{(n)})=\exp(-i\tau H_{0,n})\bigotimes_{j\in\nbbs{n}}\exp(-i\tau H_{j}),
\Eee
where the first part of the product is the time operator evolution acting on $\HH_0\otimes\HH_n$, while the second part describes the ``free'' evolution of all other modes (the modes not interacting with the system at this time). It is convenient to define, for a given Hilbert space $\HH_j$, the identity operator $\id_j$. Moreover, we write $\id_\BB=\bigotimes_{j\in\Nbb^*}\id_j$, and $\id_{\nbb{n}}=\id_0\bigotimes_{j\in\nbbs{n}}\id_j$. To keep notation as compact as possible we will omit those identities when possible. For example: $a_0\otimes\id_\BB\leftrightarrow a_0$, and identically $a_n\otimes\id_{\nbb{n}}\leftrightarrow a_n$.

\section{The Fermionic case}\label{fermion_section}
In this section we show how the fermionic case exactly maps onto the classical chaos game. Operators $a_j$ and $a^\dag_j$  now become $f_j$ and $f^\dag_j$ satisfying the anticommutation relations $\{f^\dag_j,f_i\}=\delta_{i,j}$ and $\{f_j,f_i\}=\{f^\dag_j,f^\dag_i\}=0$. Initially, the system is prepared in the vacuum state $\ket{0}_0$ such that $f_0\ket{0}_0=0$. We restrict the random variable $\gamma_j$ to take values in $\{0,1\}$. Each mode $j$ of the environment is prepared in one of the states $\ket{\beta_0}_j=\ket{0}_j$ and $\ket{\beta_1}_j=\ket{1}_j=f^\dag_j\ket{0}_j$. Here $\ket{\gamma_j}_j$ can be written explicitly in terms of the random variable $\gamma_j$: $\ket{\gamma_j}_j=(1-\gamma_j)\ket{0}_j+\gamma_j\ket{1}_j$. 
\begin{figure}
  \centering
  \includegraphics[width=1.0\linewidth]{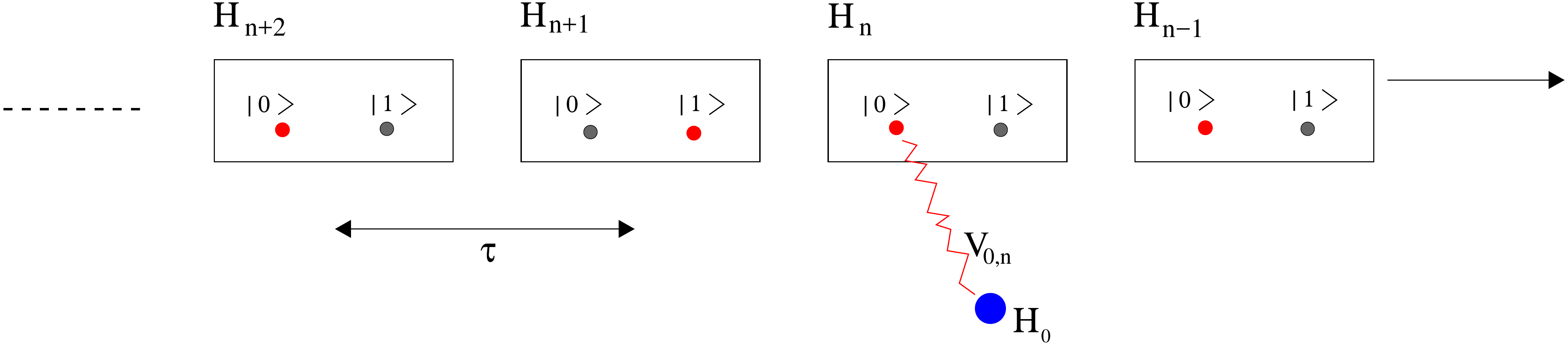}
  \caption{\label{FIG_fermion} Illustration of the quantum chaos game for fermionic particles. Each mode of the environment is prepared in one of the states $\ket{0}_n$, $\ket{1}_n$ (indicated in red).}
\end{figure}

\subsection{The action of $f_0$ on $\ket{n \tau}$}
Let us focus on the action of the annihilation operator $f_0$ on the state $\ket{n \tau}$:
\Bee
f_0\ket{n \tau}=f_0U_{(n)}(\tau)\ket{(n-1)\tau}=U_{(n)}(\tau) f_{0,(n)}(\tau) \ket{(n-1)\tau},
\label{eq_cc}
\Eee
with
\Beq
f_{0,(n)}(\tau)&=&U_{(n)}^\dag(\tau) f_0 U_{(n)}(\tau)\\
			&=&\left[\exp(i\tau H_{0,n})(f_0\otimes\id_n)\exp(-i\tau H_{0,n})\right]\otimes\id_{\nbbs{n}}.\nonumber
\Eeq
Analogously, we can define
\Beq
f_{n,(n)}(\tau)&=&U_{(n)}^\dag(\tau) f_n U_{(n)}(\tau)\\
			&=&\left[\exp(i\tau H_{0,n})(\id_0\otimes f_n)\exp(-i\tau H_{0,n})\right]\otimes\id_{\nbbs{n}}.\nonumber
\Eeq
It is not hard to show that (see appendix section \ref{fermionic_evolution}) 
\Beq\label{eq_evol_f_0}
f_{0,(n)}(\tau)=\left(e^{-i\tau \Tbf}\right)_{1,1} f_0 +\left(e^{-i\tau \Tbf}\right)_{1,2} f_n \\
\label{eq_evol_f_n}
f_{n,(n)}(\tau)=\left(e^{-i\tau \Tbf}\right)_{2,1} f_0+\left(e^{-i\tau \Tbf}\right)_{2,2} f_n,
\Eeq
where $\left(e^{-i\tau \Tbf}\right)_{i,j}$ are elements of the matrix $e^{-i\tau \Tbf}$ with $\Tbf$ given by equation \eqref{matT}. It follows from \eqref{eq_cc} that
\Bee
f_0\ket{n \tau}
=U_{(n)}(\tau) [\left(e^{-i\tau \Tbf}\right)_{1,1} f_0 +\left(e^{-i\tau \Tbf}\right)_{1,2}f_n  ] \ket{(n-1)\tau},
\label{eq20}
\Eee
so that we can check $\bra{n \tau}f_0\ket{n \tau}=0$ (see appendix section \ref{fermionic_null} for details). If one wants to measure how the environment affects the system the next natural step is to evaluate $\bra{n \tau}f_0^\dag f_0\ket{n \tau}$. 

\subsection{The average number of fermions in the system: $\NN_n$}
We define $\NN_n$ as the average number of fermions in the system at time $t=n\tau$: $\NN_n=\bra{n \tau}f_0^\dag f_0\ket{n \tau}$. In fact $\NN_n$ allows for a full description of the system. Let us remind the reader that the ensemble system-environment is prepared in a pure state $\ket{I}$ and evolves in time to be $\ket{n\tau}$. However, restricting our attention to the system only, we can define the reduced density matrix $\rho_S$, via the partial trace over the degrees of freedom of the environment, yielding $\rho_S(n\tau)={\rm Tr}_\BB\left\{\ket{n\tau}\bra{n\tau}\right\}$. Since $\bra{n \tau}f_0\ket{n \tau}=0$ we deduce that $\rho_S(n\tau)$ is Gaussian: $\rho_S(n\tau)\propto \exp(-g_nf_0^\dag f_0)$. With $\NN_n=e^{-g_n}/(1+e^{-g_n})$. Alternatively we can write $\rho_S$ under the form $\rho_S(n\tau)=(1-\NN_n)\id_0+(2\NN_n-1)f_0^\dag f_0$. We show (see appendix section \ref{fermionic_N_evolution}) that:
\Bee
\label{eqNNn}
\NN_n
=\left|\left(e^{-i\tau \Tbf}\right)_{1,1}\right|^2
\NN_{n-1}+
\left|\left(e^{-i\tau \Tbf}\right)_{1,2}\right|^2\gamma_n,
\Eee
with initial condition $\NN_0=\bra{I}f_0^\dag f_0\ket{I}=0$. Let us continue by writing $\left(e^{-i\tau T}\right)_{1,1}=e^{-i\omega\tau}\cos(\lambda\tau)$ and $\left(e^{-i\tau T}\right)_{1,2}=ie^{-i\omega\tau}\sin(\lambda\tau)$ 
and $\ww =\sin^2(\lambda\tau)$. This takes us back to the chaos game presented in section \ref{classical}:
\Bee\label{eqNNn}
\NN_n=(1-\ww)\NN_{n-1}+\ww \gamma_n.
\Eee
As a consequence, for $\lambda\tau\in]\pi/4,3\pi/4[$ (mod $2\pi$) the set of points $\{\NN_n|\forall n \in\Nbb\}$ converges towards the Cantor-Voltera fractal set. However, outside the $]\pi/4,3\pi/4[$ interval, $\NN_n$ is unrestricted in $[0,1]$. This immediately suggests a question. If playing the chaos game for $0<\ww \le 1/2$ leads to $\NN_n$ taking continuous values on $[0,1]$, how are those values distributed? Can we evaluate the density of points in the interval?  
\subsection{Point density for $0<\ww \le1/2$}
Since we have already understood the process leading to fractal sets when considering $1/2<\ww <1$, we here focus on the point density for $\NN_n$ when $0<\ww \le1/2$. The ergodic theorem for iterated maps \cite{Elton_87,Forte_98} allows us to reach the density distribution from two different points of view.
\begin{enumerate}
\item Let us first define $T_{n}(x,y)$ as the total number of points $\NN_j$, generated after $n$ iterations, and such that $x\le \NN_j<y$:
\begin{equation}
T_{n}(x,y)=\sum_{j=0}^n\Pi_{x,y}(\NN_j),
\end{equation}
with
\begin{equation}
\Pi_{x,y}(\NN_j)=
\left\{
\begin{array}{ccc}
1 & {\rm if} & x\le \NN_j<y\\
0 & &�{\rm otherwise}.
\end{array}
\right.
\end{equation}
We divide $[0,1]$ in $K$ intervals and write $\sum_{j=0}^{K-1}T_{n}(j/K,(j+1)/K)=n+1$ taking the limit $K\rightarrow\infty$. This allows us to define the density $\sigma_n(x)$ as $\sigma_n(x)=\lim_{K\rightarrow\infty}\frac{K}{n+1}T_{n}(j/K,(j+1)/K)$ satisfying $\int_0^1dx\sigma_n(x)=1$. In particular we are interested in $\sigma^*(x)=\lim_{n\rightarrow\infty}\sigma_n(x)$.

\item Assume that the starting point of the chaos game $x_0$ is randomly distributed according to the density distribution $\eta_{0}(x)$. We define $\eta_{n}(x)$ the distribution of points after $n$ iterations. One can easily show that the latter quantity must obey the following equation:
\Bee
\label{density_1D}
\eta_{n+1}(x)=\frac{1}{2}\frac{1}{1-\ww }\left[\eta_n\left(\frac{x}{1-\ww }\right)+\eta_n\left(\frac{x-\ww }{1-\ww }\right)\right].
\Eee
Note that if $\eta_0(x)$ is such that $\eta_0(x)=\eta_0(1-x)$ then $\eta_n(x)=\eta_n(1-x)$ $\forall n\in\Nbb$. Once again, we write $\eta^*(x)=\lim_{n\rightarrow\infty}\eta_n(x)$. The latter equation allows us to write for any function $h(x)$:
$
2\langle h(x)\rangle=\langle h(x(1-\ww))\rangle+\langle h(x(1-\ww)+\ww )\rangle
$. Choosing $h(x)=x^n$ gives another derivation of the recursively defined moment formula already presented in \cite{Lad_92,Hosking_94,Grabner_96} 
\begin{eqnarray}\label{moment_eq}
\langle x^n\rangle=\frac{1}{2(1-(1-\ww)^n)}\sum_{i=0}^{n-1}C^n_i(1-\ww)^i\ww^{n-i}\langle x^i\rangle.
\end{eqnarray}
\end{enumerate}
{
Alternatively, we can think of the chaos game as a random walk by rewritting the chaos game equation as $x_{n+1}=x_{n}+\Delta_{\pm}(x_n)$, where $\Delta_-(x_n)=-\ww x_n$ and $\Delta_+(x_n)=\ww(1-x_n)$ are the lengths of the left and right steps from position $x_n$. In particular we see that $|\Delta_+|+|\Delta_-|=\ww$. In the statistical physics community, random walks with variable size steps have been considered in \cite{Krapisvsky_03,Turban_10}. To make the link between the chaos game and those studies more explicit, we start from the chaos game equation \eqref{chaos_equation} (using $x_0=0$) and define $\sigma_j=2c_{n-j-1}-1$ as spin variables (taking values in $\{-1,1\}$). We show by induction
\Bee\label{s_sigma}
x_n=\frac{1-(1-\ww)^n}{2}+\frac{\ww}{2}\sum_{j=0}^{n-1}(1-\ww)^j\sigma_j,
\Eee
with $\langle\sigma_i\rangle=0$ and $\langle\sigma_i\sigma_j\rangle=\delta_{i,j}$. From the previous equation, we see that the distribution of the variable $x_n$ is determined by the distribution of the random variable $\Theta_n=\sum_{j=0}^{n-1}\bar\ww^n\sigma_n$, with $\bar\ww=1-\ww$. In the mathematical community, the distribution of $\Theta$ is known as Bernoulli convolutions \cite{Kershner_35,JESSEN_35,Varju_16,Peres_00,Erdos_39,Wintner_35,Solomyak_95,Erdos_40}. In 1935 Jessen and Wintner \cite{JESSEN_35} proved that for $0<\bar\ww<1$ the distribution of $\Theta$ is either absolutely continuous or singular with respect to Lebesgue measure. Since then the challenge has been to identify the set of $\bar\ww$ values for which the distribution is absolutely continuous or singular. Amongst the many noticeable results (see reviews \cite{Varju_16,Peres_00}), P. Erdos \cite{Erdos_39} showed that the density is singular for all $\bar\ww\in ]1/2,1[$ such that $1/\bar\ww$ is a Pisot number. No other than reciprocal Pisot numbers are known to be associated to a singular distribution \cite{Peres_00}. On the other hand, the density is known to be absolutely continuous for $\bar\ww=2^{-1/m}$ \cite{Wintner_35}. An exact expression of the probability distribution (in real space) for $\bar\ww=2^{-1/m}$ with $m=1,2,3$ can be found in \cite{Krapisvsky_03}. It is continuous and piecewise polynomial of maximum order $m-1$. Moreover, Solomyak \cite{Solomyak_95} proved that the density is absolutely continuous almost everywhere for $\bar\ww\in]1/2,1[$. In addition, P. Erdos \cite{Erdos_40} showed that the density is absolutely continuous for $\bar\ww$ sufficiently close to one. However, no explicit bound for the neighbourhood of one is given. 
}

\begin{figure}[h]
  \centering
  \includegraphics[width=0.495\linewidth]{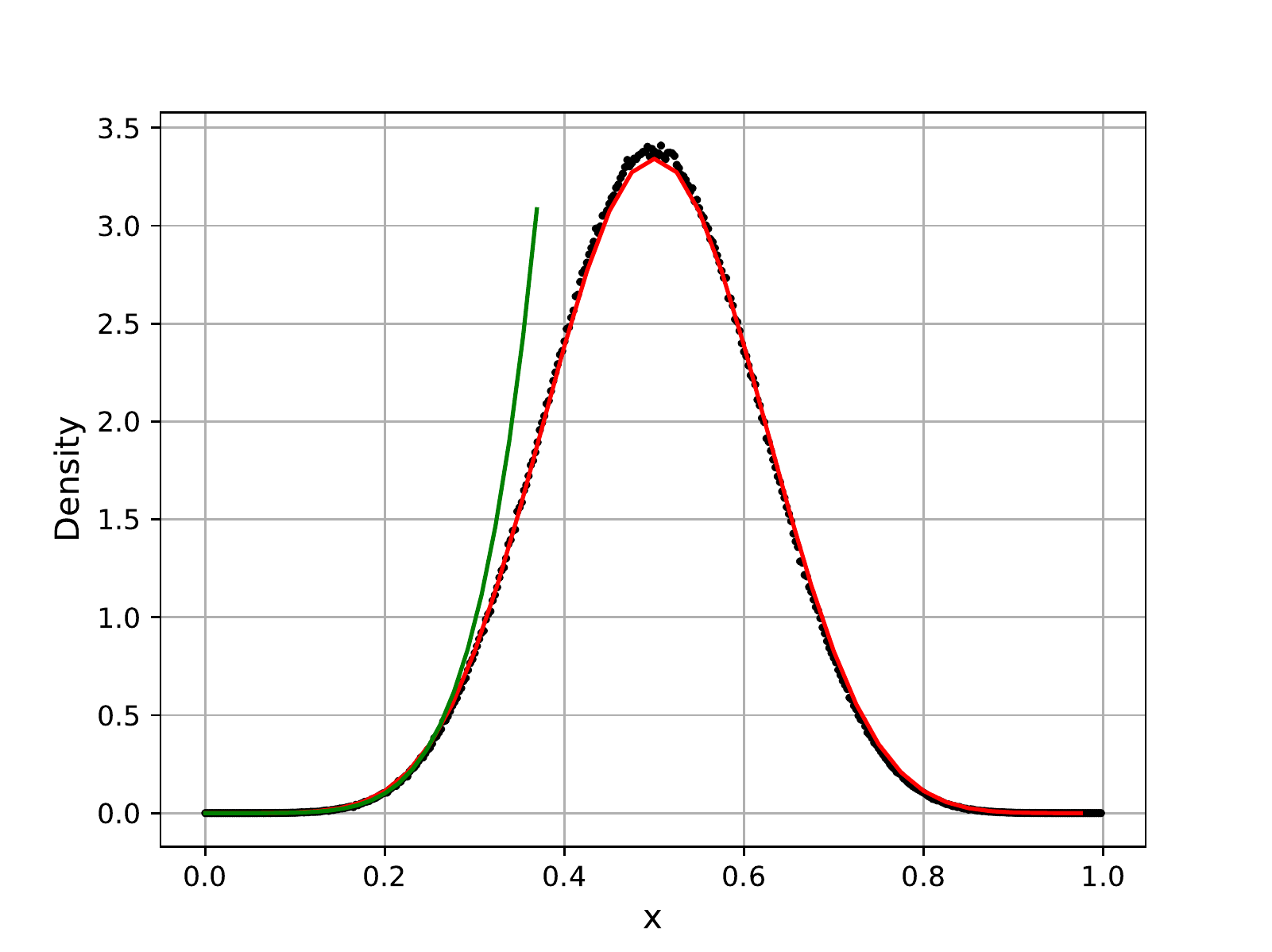}
  \includegraphics[width=0.495\linewidth]{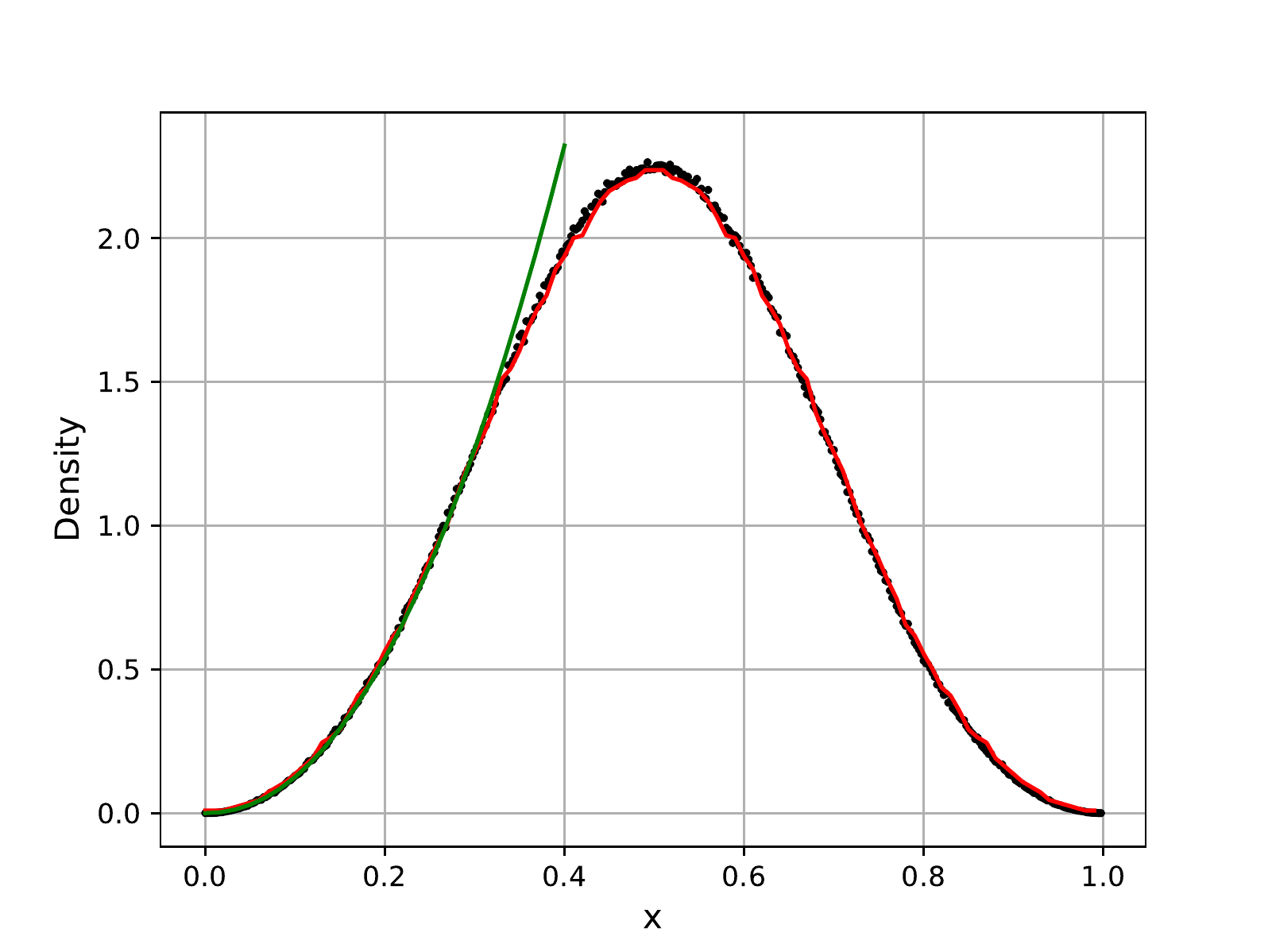}\\
  \includegraphics[width=0.495\linewidth]{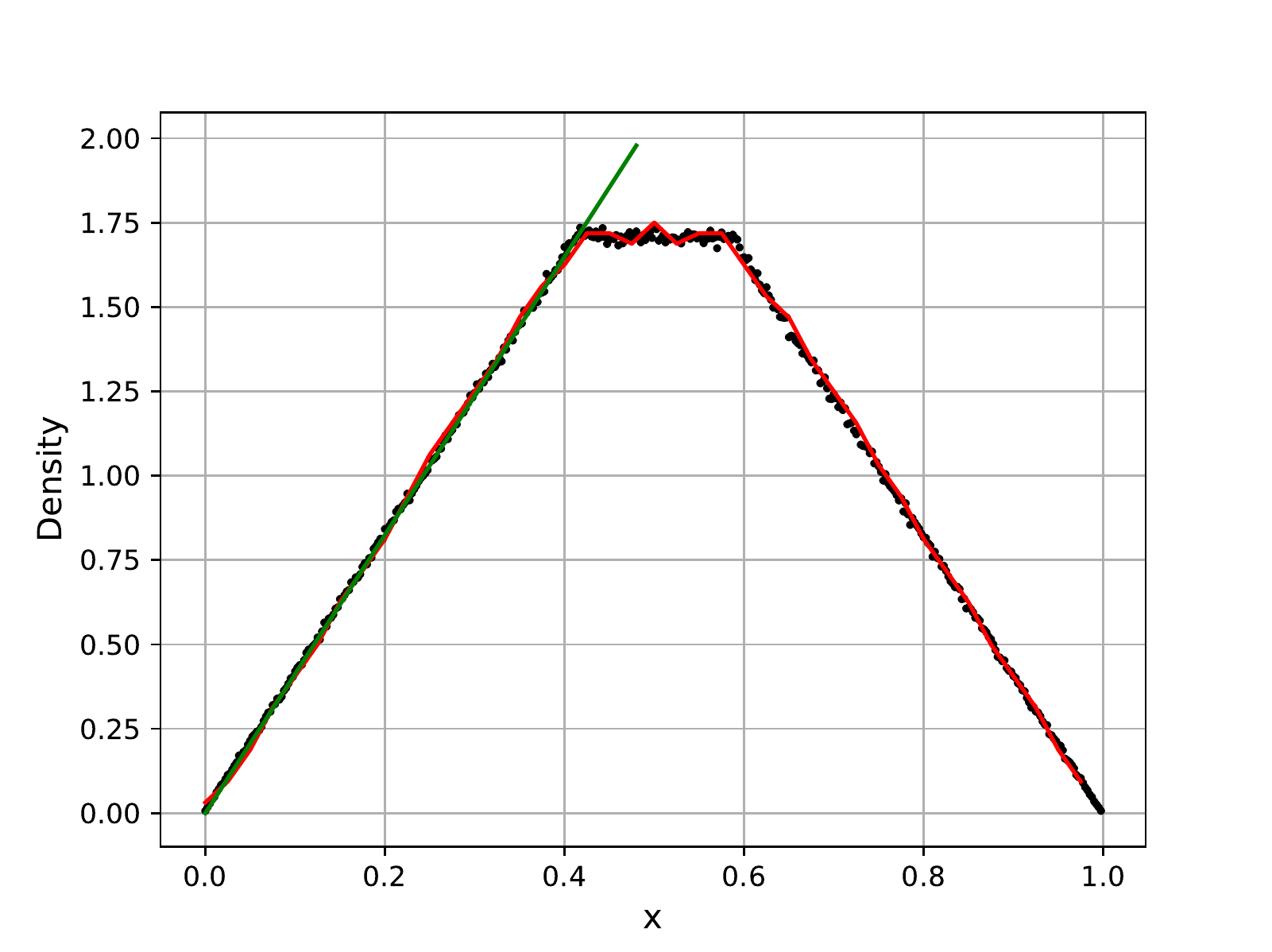}
  \includegraphics[width=0.495\linewidth]{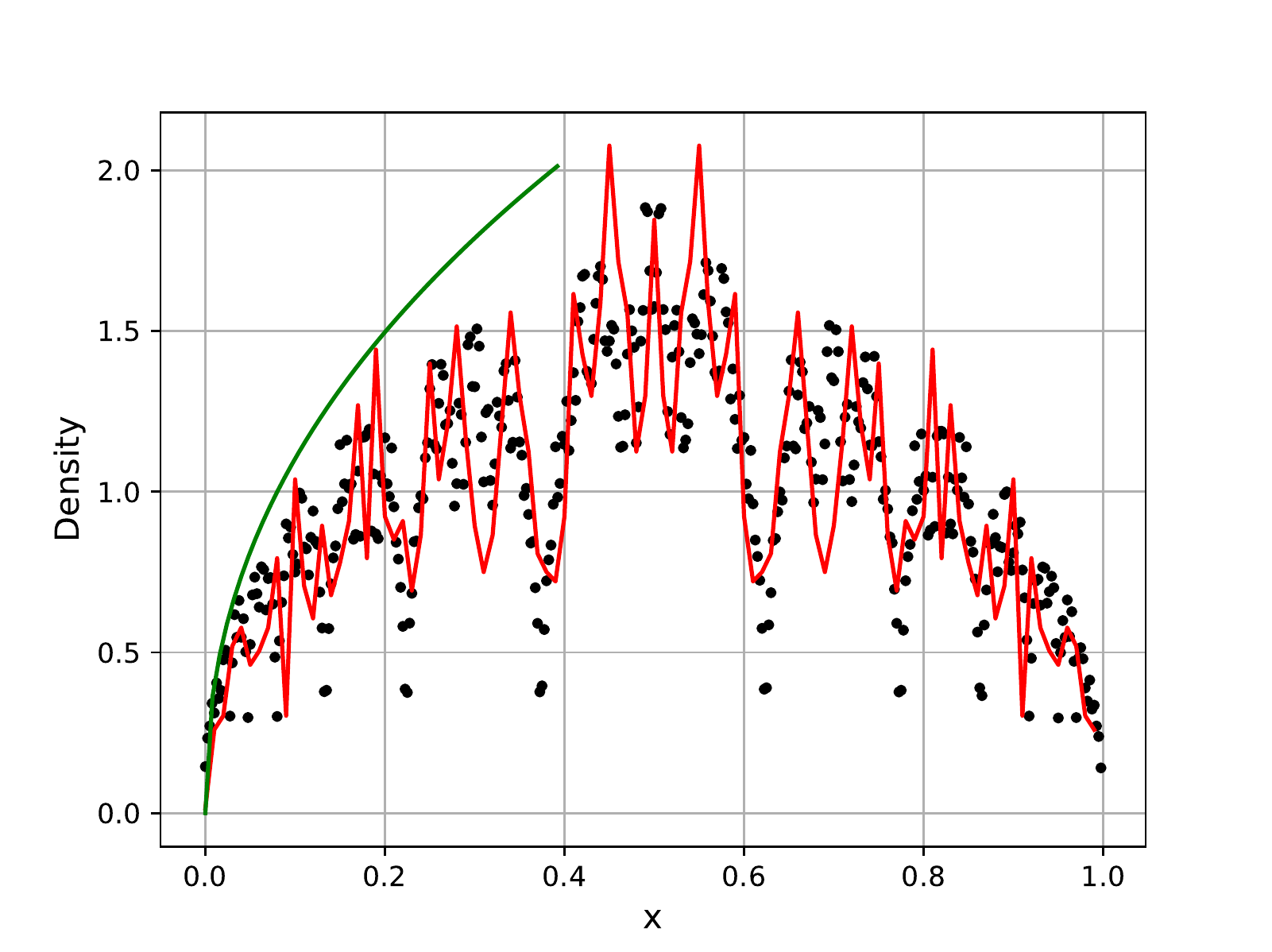}
  \caption{\label{density} Comparison of density plots $\sigma_n(x)$ and $\eta_m(x)$. In all four plots, the dots represent the distribution $\sigma_n(x)$ generated by numerical simulation of the chaos game for $n=5\times 10^6$ iterations. The red lines are associated to the distribution $\eta_{m}(x)$ also evaluated numerically. The green lines are fits using the ansatz $Ax^\alpha$. In the top left corner, we present results for $\ww=0.1$. The distribution $\sigma_n(x)$ is compared to $\eta_{m}(x)$ for $m=25$. In the top right corner, we present results for $\ww=0.2$. The distribution $\sigma_n(x)$ is compared to $\eta_{m}(x)$ for $m=10$. In the bottom left corner, we present results for $\ww=\ww_c=1-1/\sqrt{2}$. The distribution $\sigma_n(x)$ is compared to $\eta_{m}(x)$ for $m=10$. The green line is here ${(1+\sqrt{2})^2}x/{\sqrt{2}}$. In the bottom right corner, we present results for $\ww=1-1/\phi$ where $\phi=(1+\sqrt{5})/2$ is the golden ratio. The distribution $\sigma_n(x)$ is compared to $\eta_{m}(x)$ for $m=20$. The ansatz for small $x$ breaks down, only giving an envelope to the distribution.
  }
\end{figure}

\begin{figure}
  \centering
  \includegraphics[width=0.495\linewidth]{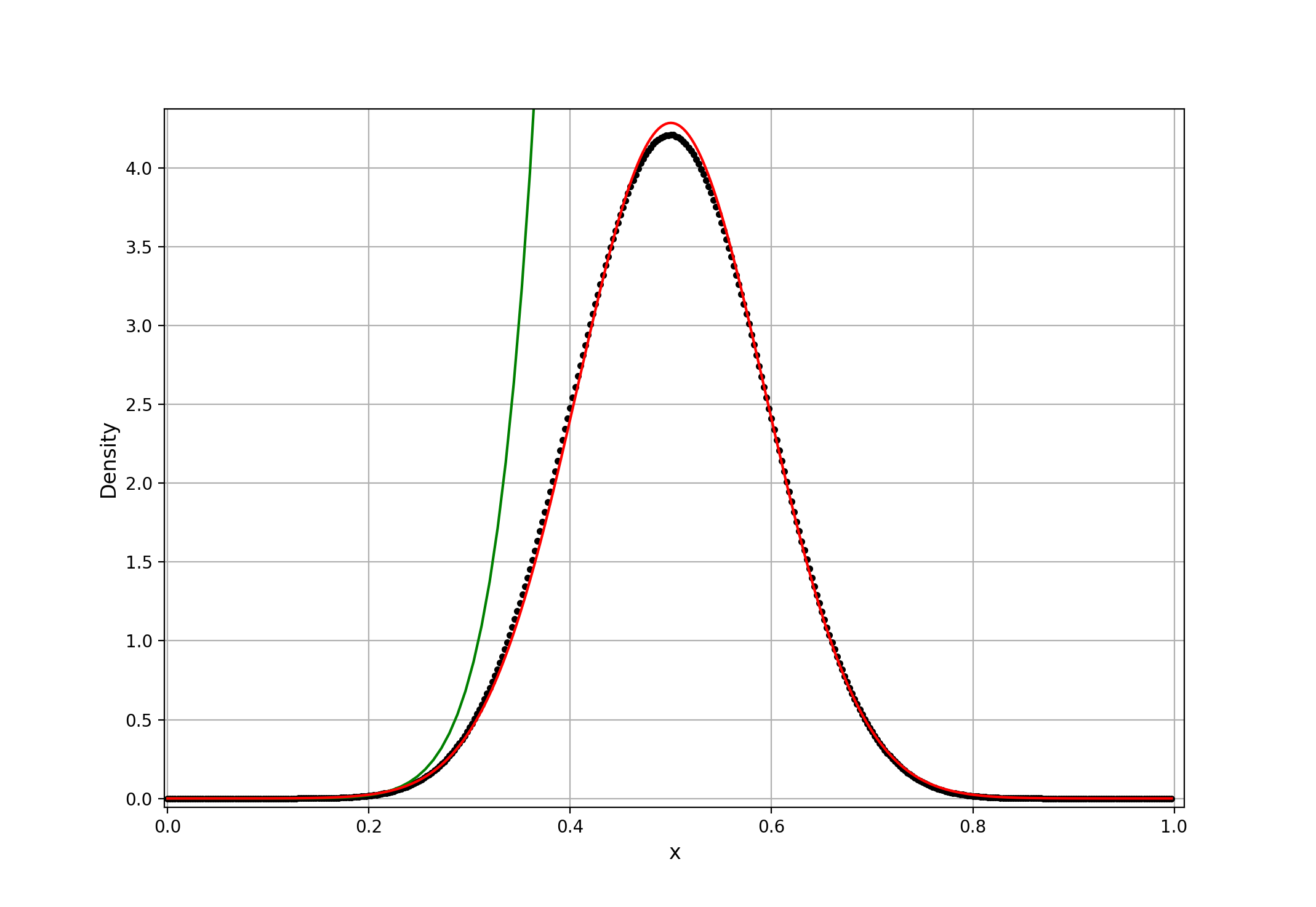}
  \includegraphics[width=0.495\linewidth]{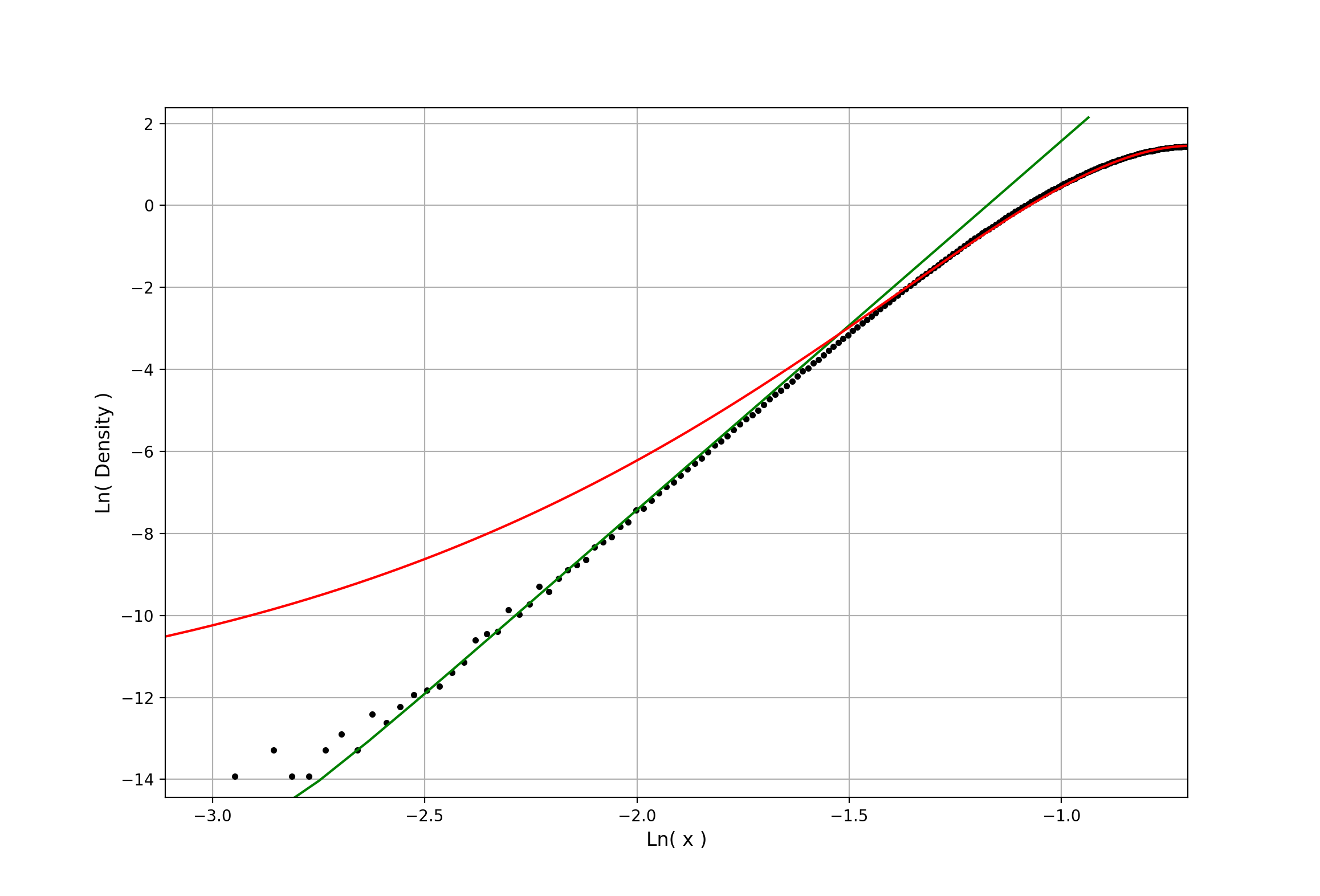}\\
  \caption{\label{density_2}{ For $\ww=1-2^{-1/10}$, comparison of density plots $\sigma_n(x)$ (black dots - generated by $n=5\times10^8$ iterations) with the Gaussian approximation (red - with mean $\langle x\rangle=1/2$ and variance $(\ww/2)^2/[1-(1-\ww)^2]$) and the small $x$-approximation (green - $\eta^*(x)\simeq Ax^9$). Linear-Linear plot on the left and Log-Log plot on the right. The Gaussian approximation is expected to give better results as $\ww$ tends to zero. In particular, the Gaussian approximation fails to described the density for small $x$, where $\rho^*(x)\propto x^9$.}
}
\end{figure}

In figure \ref{density}, we present numerical simulations for $\sigma_n(x)$ and numerical calculations for $\eta_m(x)$. Numerical results clearly confirms that both $\eta_m$ and $\sigma_n$ converge towards the same stationary distribution. We observe that the shape of the density function is strongly dependent on the $\ww$ value. {Up to a translation and dilatation, the plots presented here are similar to the ones in \cite{Krapisvsky_03,Turban_10}. For $\ww=1-1/\phi$ with $\phi=(1+\sqrt{5})/2$  (the golden ratio), a detailed analysis of the sefsimilar distribution was presented in \cite{Krapisvsky_03}.} For $\ww =1/2$ one trivially verify that $\eta^*(x)=1$ (for all $x\in]0,1[$) and for $\ww =\ww _c=1-1/\sqrt{2}$ (see figure \ref{density}) one has
\begin{equation}
\label{eta_for_wc}
\eta^*(x)=
\frac{(1+\sqrt{2})^2}{\sqrt{2}}\left\{
\begin{array}{cc}
x & 0\le x < 1/(1+\sqrt{2})\\
\frac{1}{1+\sqrt{2}} & 1/(1+\sqrt{2})< x \le \sqrt{2}/(1+\sqrt{2})\\
1-x & \sqrt{2}/(1+\sqrt{2}) < x \le 1.
\end{array}
\right.
\end{equation} 
{For small $\ww$ values, one can approximate $\eta^*(x)$ (around $x=1/2$) with the Gaussian distribution characterised by mean $\moyenne{x}=1/2$ and variance ${\rm Var}=(\ww/2)^2/[1-(1-\ww)^2]$ (see figure \ref{density_2}). However, we remind the reader that $\eta^*$ must satisfy $\eta^*(0)=\eta^*(1)=0$. It follows that the Gaussian approximation fails for $x$ close to the boundaries.} For small $x$, one propose to estimate the behaviour of $\eta^*$ using equation \eqref{density_1D} and $\eta^*(x)\propto x^\alpha$ as an ansatz. Indeed, for $x<\ww $ one has $\eta^*\left(\frac{x-\ww }{1-\ww }\right)=0$ so that we are left with $2(1-\ww)\eta^*(x)=\eta^*\left({x}/{(1-\ww)}\right)$, which leads to
\Bee
\alpha(\ww )=-1-\frac{\ln(2)}{\ln(1-\ww)}.
\Eee
In particular, we recover $\alpha=1$ for $\ww =\ww _c$ in agreement with \eqref{eta_for_wc}. {Inverting the last equation allows us to find $\ww$ values for which one can expect a behaviour of the form $x^\alpha$. In particular for $\alpha=m$ integer, one recover all $\ww$ values given by $1-\ww=2^{-1/(m+1)}$ for which the density is known to be absolutely continuous \cite{Wintner_35}.} One should note that $\ww =\ww _c$ separates density distributions characterised by $\alpha<1$ (for $\ww >\ww _c$) and $\alpha>1$ (for $\ww <\ww _c$).

\subsection{Entanglement Entropy}
\begin{figure}
  \centering
  \includegraphics[width=0.8\linewidth]{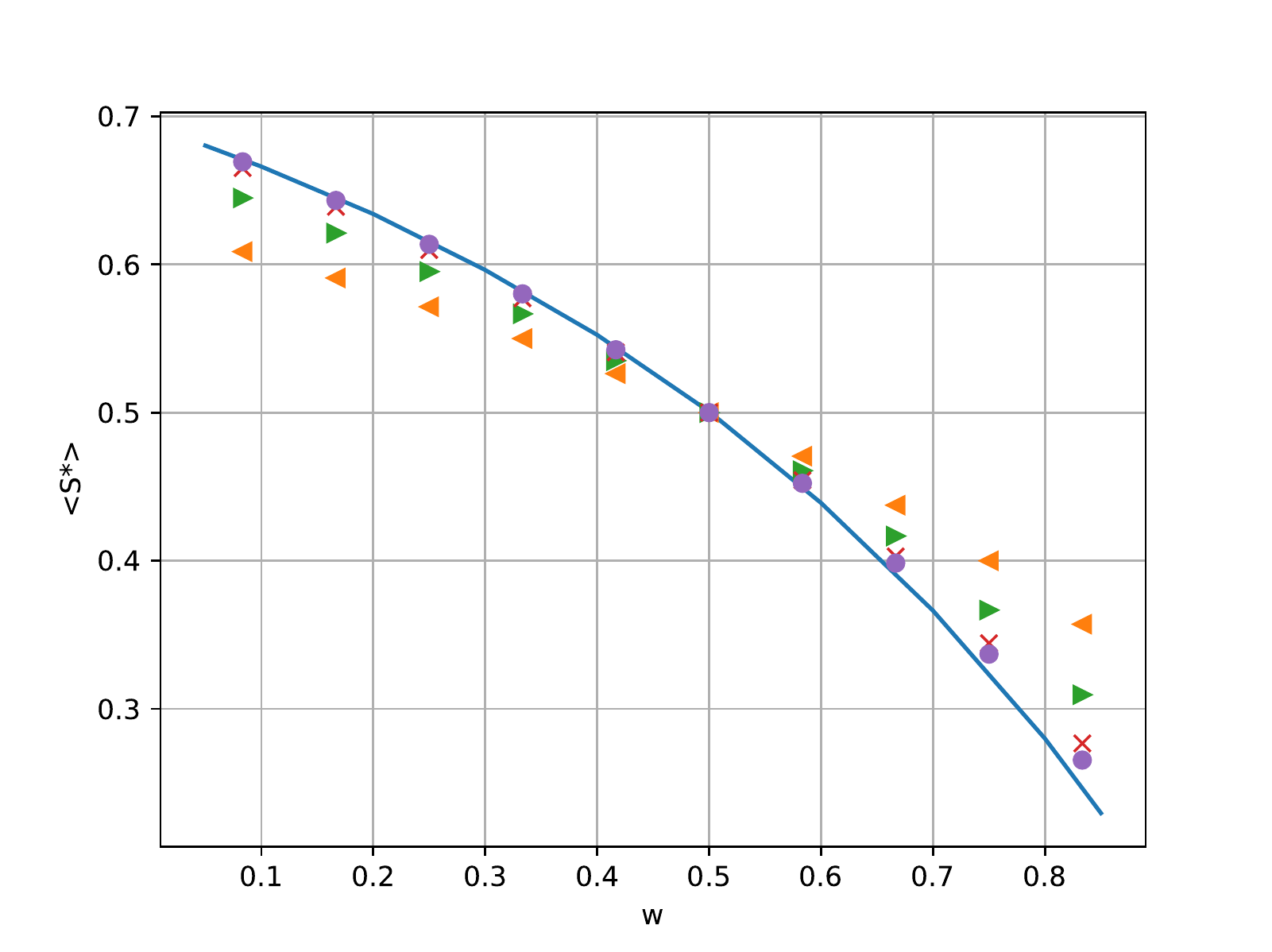}\\
  \caption{\label{Entropy} The entanglement entropy $\langle S^*\rangle$ as a function of $\ww$. The line is associated to the time average generated via numerical simulation of the chaos game. The symbols are results from approximation \eqref{approx_S} with truncation order $K=2,3,5$ and $7$ (respectively left triangle, right triangle, cross and circle). {One note that the approximation converges faster towards $\langle S^*\rangle$ when $\ww<1/2$.}
  \vspace{2cm}\mbox{ }}
\end{figure}

Let us once again insist on the fact that the system is described by a mixed state ensemble (the density matrix $\rho_S$), while the ensemble system-environment is in a pure state ($\ket{n\tau}$). This is in fact a direct signature of the entanglement which links the system and the bath. Starting from a factorised pure state $\ket{I}=\ket{0}_0\otimes\ket{\Gamma}_\BB$ the dynamics naturally entangles the system and bath states. In other words, at a later time $t=n\tau$ it is not possible to write $\ket{n\tau}$ as a direct product of the form $\ket{\phi}_0\otimes\ket{\psi}_\BB$. {One should mention that, in the mathematical community, work on the entropy for Bernoulli convolutions have been considered in \cite{Garsia_63,Alexander_91}. However, in this context, the focus is quite different. Indeed, those studies are aimed to analyse the entropy associated to the random variable $\Theta_n=\sum_j^n\bar\ww^j\sigma_j$, defined in \eqref{s_sigma}. In our study we focus the average number of fermions $\NN_n$ which if related to $\Theta_n$ (via \eqref{s_sigma}) is limited to only taking values $0$ or $1$ with probabilities $1-\NN_n$ and $\NN_n$.} One measure of entanglement is given by the Von Neumann entropy $S_n=-{\rm Tr}\left\{\rho_S(n\tau)\ln[\rho_S(n\tau)]\right\}$. In a sense it measures how far the density matrix of the sytem is to describing a pure state. One should mention that entanglement is a purely quantum property which has been the focus of an enormous number of studies. It plays a key role in quantum engineering and information processing \cite{Rabitz_00,Chu_02,Dowling_03,Reichle_06,Castelvecchi_17,Aleksandra_18}. In relation to the repeated interaction process entropy production and entanglement have been studied in \cite{Benoist_18} and \cite{Attal_14,Wendenbaum_15} respectively. At a given time $n\tau$, the entropy is $S_n=-(1-\NN_n)\ln(1-\NN_n)-\NN_n\ln\NN_n$. Focusing on the average entanglement entropy 
we performe numerical calculations of $\langle S^*\rangle=\lim_{n\rightarrow\infty}\frac{1}{n}\sum_{j=1}^nS_n$ for various values of $\ww $ (see figure \ref{Entropy}). Alternatively (using the property $\eta^*(x)=\eta^*(1-x)$) we can write
\Bee\label{eqS}
\langle S^*\rangle=-2\int_0^1 dxx\ln(x)\eta^*(x).
\Eee
One observes that the entanglement $S_n$ reaches its maximum $S_{max}=\ln(2)$ for $\NN_n=1/2$. Moreover, for $\ww >1/2$ the fractal structure does not allow $\NN_n$ to take values around $1/2$ in the interval $]1-\ww ,\ww [$. It follows that highly entangled states are forbidden by the dynamics. In contrast, for $\ww <1/2$ highly entangled states are now accessible and even most probable. At this stage $\langle S^*\rangle$ can be evaluate exactly for $\ww =1/2$ and $\ww =\ww _c$ only, as in both case, $\eta^*(x)$ is known. For all other $\ww$ value, we propose the following strategy. Starting from equation \eqref{eqS}, changing the variable $x\rightarrow1-x$ and using the Taylor expansion $\ln(1-x)=-\sum_{n=1}^\infty x^n/n$ one can express $\langle S^*\rangle$ as a sum over all moments $\langle x^n\rangle$ (which recursively can be calculated exactly using \eqref{moment_eq}):
\Bee\label{eq26}
\frac{\langle S^*\rangle}{2}=\sum_{n=1}^\infty \frac{\langle x^n\rangle-\langle x^{n+1}\rangle}{n}=\langle x\rangle-\sum_{n=2}^\infty\frac{\langle x^n\rangle}{n(n-1)}.
\Eee
Truncating the calculation to a given order $K$, we define $A_K=\sum_{n=1}^{K-1} ({\langle x^n\rangle-\langle x^{n+1}\rangle})/{n}$ and $B_K=\langle x\rangle-\sum_{n=2}^K{\langle x^n\rangle}/{n(n-1)}$. Since $A_K<\langle S^*\rangle/2<B_K$, we propose approximating $\langle S^*\rangle$ using $C_K=(A_K+B_K)/2$, leading to
\begin{eqnarray}\label{approx_S}
\langle S^*\rangle \simeq 1-2\sum_{n=2}^{K-1}\frac{\langle x^n\rangle}{n(n-1)}-\langle x^K \rangle\frac{K+1}{K(K-1)}.
\end{eqnarray}
The comparison between $\langle S^*\rangle$ obtained under numerical simulation of the chaos game with the approximation \eqref{approx_S} for order $K=2,3$ and $5$ is presented in figure \ref{Entropy}. {One should note that the truncated approximation converges faster towards $\langle S^*\rangle$ when $\ww<1/2$.}

\section{The Bosonic case}\label{boson_section}
When considering the bosonic scenario we write $b_j$ and $b^\dag_j$ for the anihilation and creation operators. As we will see below, by preparing the system in a coherent state $\ket{\chi_0}_0$, we obtain a situation in which the Barnsley's $2d-$chaos game arises. A coherent state is an eigenvector of the annihilation operator $b_0$ such that $b_0\ket{\chi_0}_0=\chi_0\ket{\chi_0}_0$ with $\chi_0\in\Cbb$:
\Bee
\ket{\chi_0}_0=e^{-|\chi_0|^2/2}\sum_{n=0}^\infty\frac{\chi_0^n}{\sqrt{n!}}\ket{n}_0,
\Eee
where $\ket{n}_0$ is the state populated by $n$ particles: $b^\dag_0b_0\ket{n}_0=n\ket{n}_0$. We let the random variables $\gamma_j$ take values in $\{0,1,2,..., M-1\}$ and define $M$ coherent states $\ket{\beta_j}$, for $j\in\{0,1,2,\hdots,M-1\}$ parametrised by complex numbers $\beta_j$ which we can choose freely. Initially, every mode of the environment is prepared, at random, in one of the coherent states $\ket{\beta_j}$ with equal probability. The full initial state is $\ket{I}=\ket{\chi_0}_0\otimes\ket{\Gamma}_\BB$, with $\ket{\Gamma}=\bigotimes_{j\in\Nbb^*}\ket{\gamma_j}_j$ and $\ket{\gamma_j}_j=\sum_{k=0}^{M-1}\delta_{\gamma_j,k}\ket{\beta_k}_j$. Since $\ket{\gamma_j}_j$ is one of the possible coherent state it is convenient to write $b_j\ket{\gamma_j}_j=\beta_{\gamma_j}\ket{\gamma_j}_j$.
\begin{figure}
  \centering
   \includegraphics[width=1.1\linewidth]{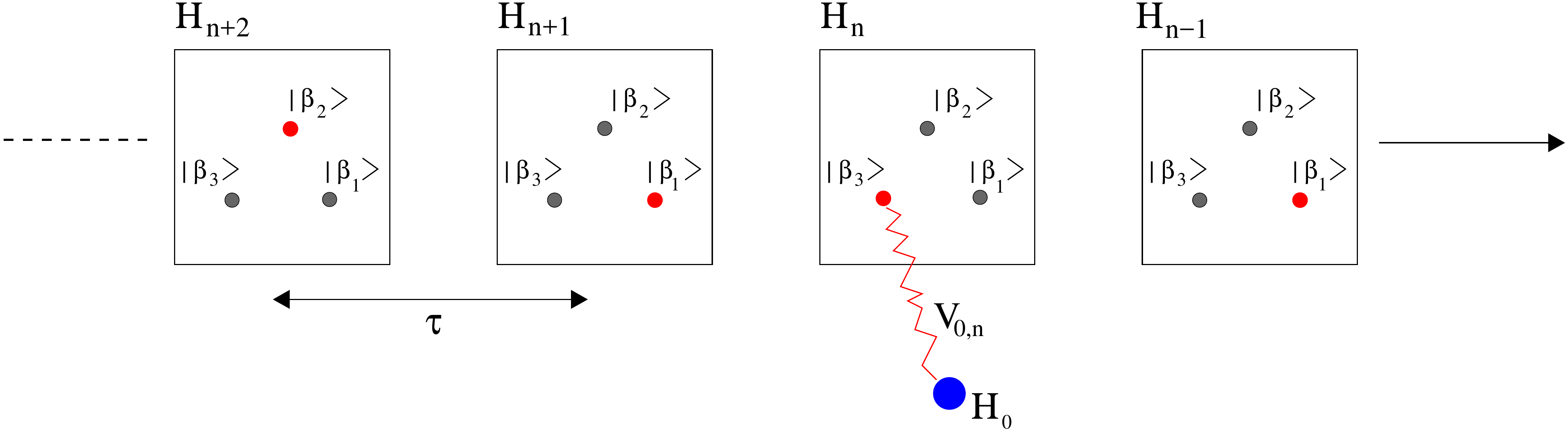}\\
  \caption{\label{FIG_boson} Illustration of the quantum chaos game for bosonic particles. Each mode is prepared in one of three coherent states $\ket{\beta_0}_n$, $\ket{\beta_1}_n$ and $\ket{\beta_2}_n$ (indicated in red).}
\end{figure}

\subsection{The action of $b_0$ on $\ket{n \tau}$}
As before, we focus on  $b_0\ket{n \tau}$:
\Bee
b_0\ket{n \tau}
			 =U_{(n)}(\tau) b_{0,(n)}(\tau) \ket{(n-1)\tau}
\label{eq_aa}
\Eee
with $b_{0,(n)}(\tau)=U_{(n)}^\dag(\tau) b_{0} U_{(n)}(\tau)$ and $b_{n,(n)}(\tau)=U_{(n)}^\dag(\tau) b_{n} U_{(n)}(\tau)$. A derivation similar to the one presented for fermions leads to $
b_{0,(n)}(\tau)=\left(e^{-i\tau \Tbf}\right)_{1,1}b_{0} +\left(e^{-i\tau \Tbf}\right)_{1,2}b_n$ and
$b_{n,(n)}(\tau)=\left(e^{-i\tau \Tbf}\right)_{2,1} b_0+\left(e^{-i\tau \Tbf}\right)_{2,2}b_n$. It follows that
\Bee
b_0\ket{n \tau}= \left(e^{-i\tau T}\right)_{1,1} U_{(n)}(\tau)  b_0 \ket{(n-1)\tau}
+\left(e^{-i\tau T}\right)_{1,2} U_{(n)}(\tau) b_n \ket{(n-1)\tau}.
\Eee
The second term of the sum can be written as 
$e^{-i\omega(n-1)\tau} \beta_{\gamma_n} \ket{(n-1)\tau}$. Starting with $b_0\ket{I}=\chi_0\ket{I}$, we show by induction that $\ket{n \tau}$ is an eigenstate of $b_0$:
$b_0\ket{n \tau}= \chi_n\ket{n\tau}$, with 
\Bee
\label{eq_chi_evol}
\chi_{n+1}=\cos(\lambda\tau)e^{-i\omega\tau}\chi_{n}
+ie^{-i\omega (n+1)\tau}\sin(\lambda\tau)\beta_{\gamma_{n+1}}.
\Eee

\subsection{The orbits of $\chi_{n}$}
\begin{figure}
  \centering
	\includegraphics[width=0.495\linewidth]{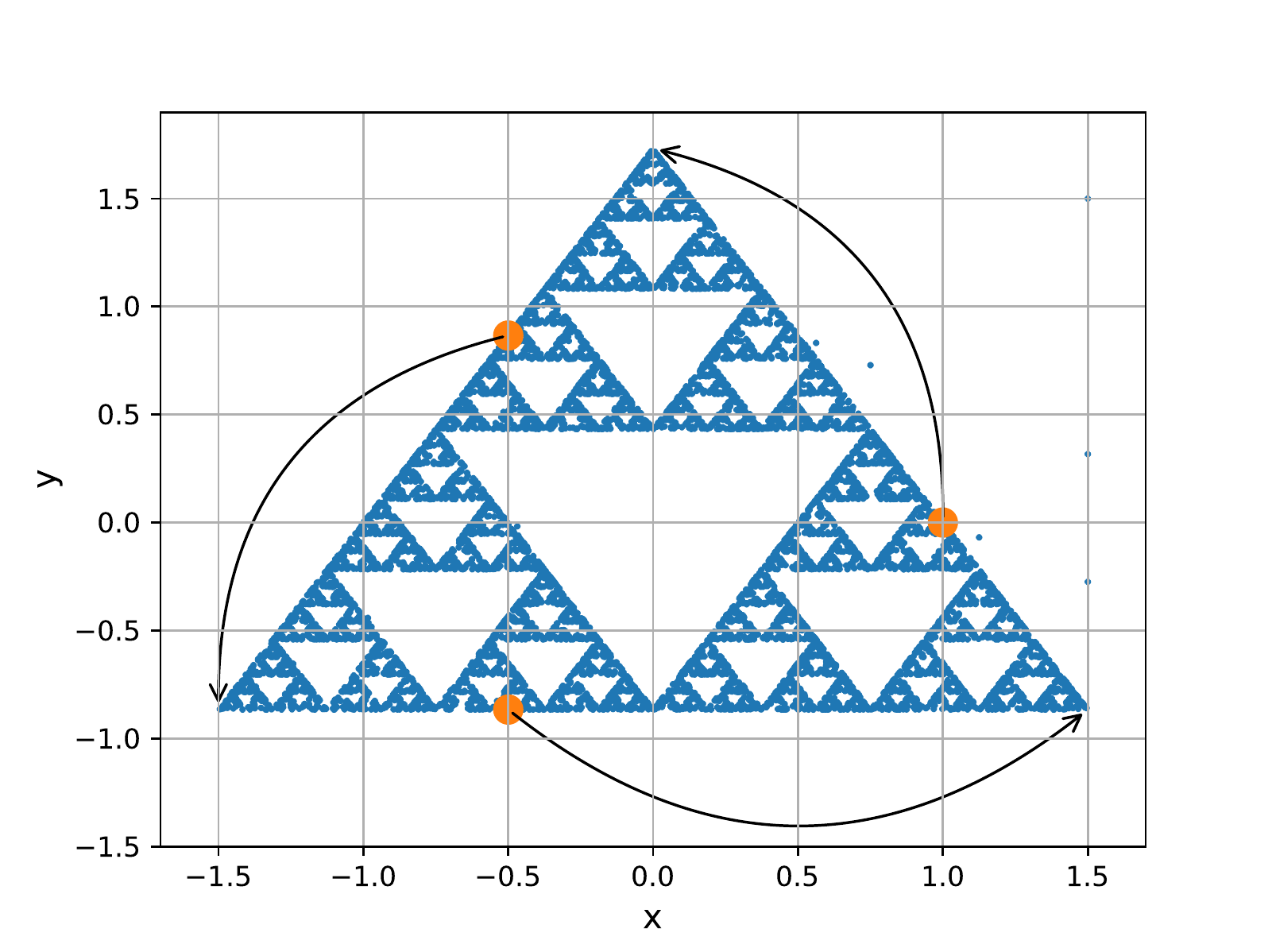}
	\includegraphics[width=0.495\linewidth]{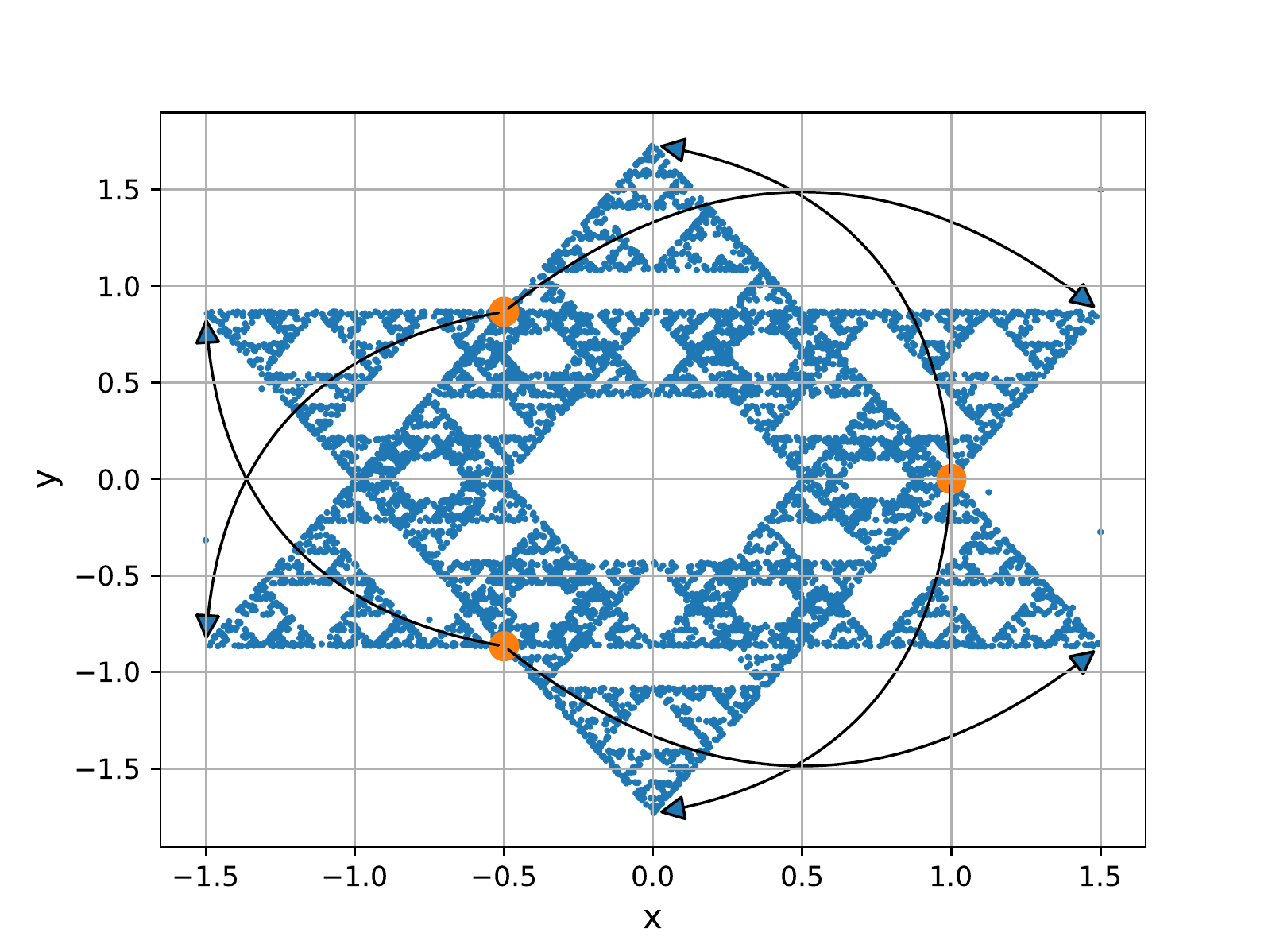}\\
  \caption{\label{Quantum_shapes_1} Simulation of Quantum chaos game for bosonic particles. In both cases $\tau=1$ and $\lambda=\pi/3$. The coherent states are indicated by the three larger circles for coordinates $\beta_j=e^{i\theta_j}$ with $\theta_j=2\pi j/3$ and $j=0,1,2$. Black arrow takes us from every $\beta_j$ to its transformations $\tilde\beta_{n,j}$. On the left $\omega=2\pi$, on the right $\omega=\pi$. A few points outside the fractal structure can be seen. These points correspond to the $\chi_n$ values for small $n$ values, as the system converge towards the attractor from its initial state.
}
\end{figure}

\begin{figure}
  \centering
	\includegraphics[width=0.495\linewidth]{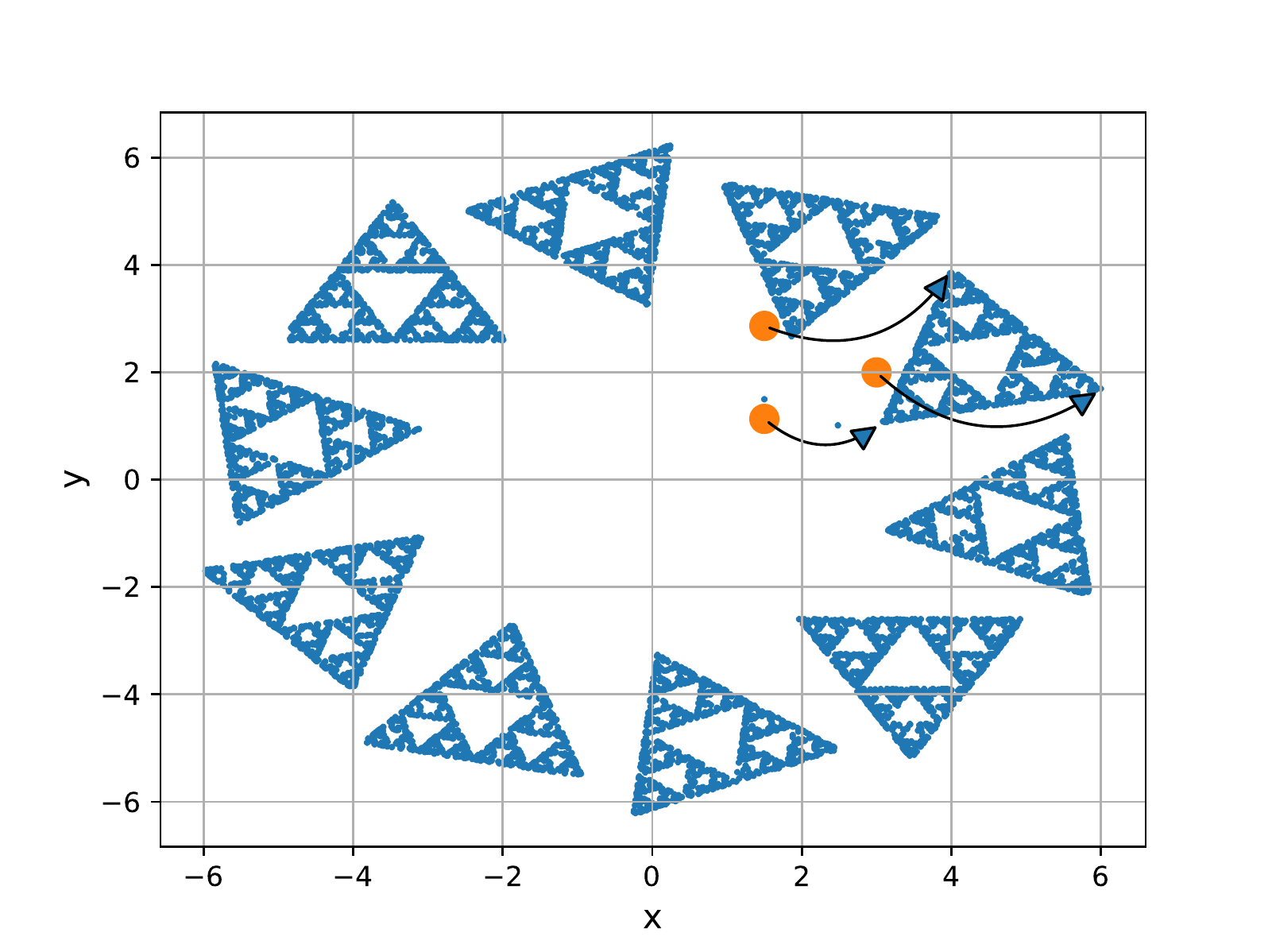}
	\includegraphics[width=0.495\linewidth]{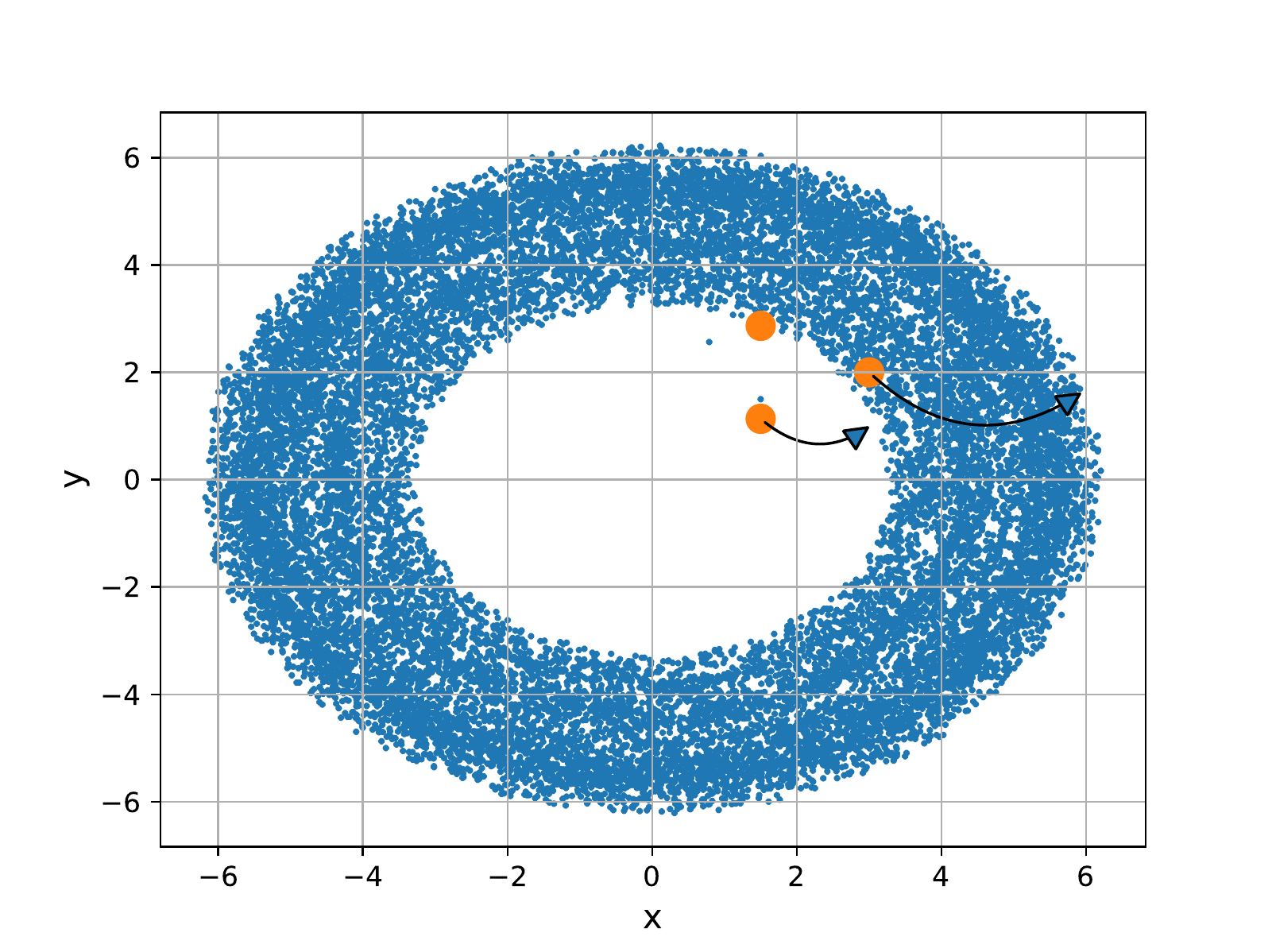}
  \caption{\label{Quantum_shapes_2} Simulation of quantum chaos game for bosonic particles. In both cases $\tau=1$ and $\lambda=\pi/3$. The coherent states are indicated by the three larger circles of coordinates $\beta_j=e^{i\theta_j}+2(1+i) $ with $\theta_j=2\pi j/3$ and $j=0,1,2$. On the left $\omega=3\pi/5$. The black arrow takes us from every $\beta_j$ to one of the transformations $\tilde\beta_{n,j}$. On the right $\omega=1$. We only indicate two arrows pointing to the $\tilde\beta$ located at the edge of the orbit. A few points outside the fractal structure can be seen. These points correspond to the $\chi_n$ values for small $n$ values, as the system converge towards the attractor from its initial state.
  \vspace{3cm}\mbox{ }}
\end{figure}

\begin{figure}
  \centering
	\includegraphics[width=0.495\linewidth]{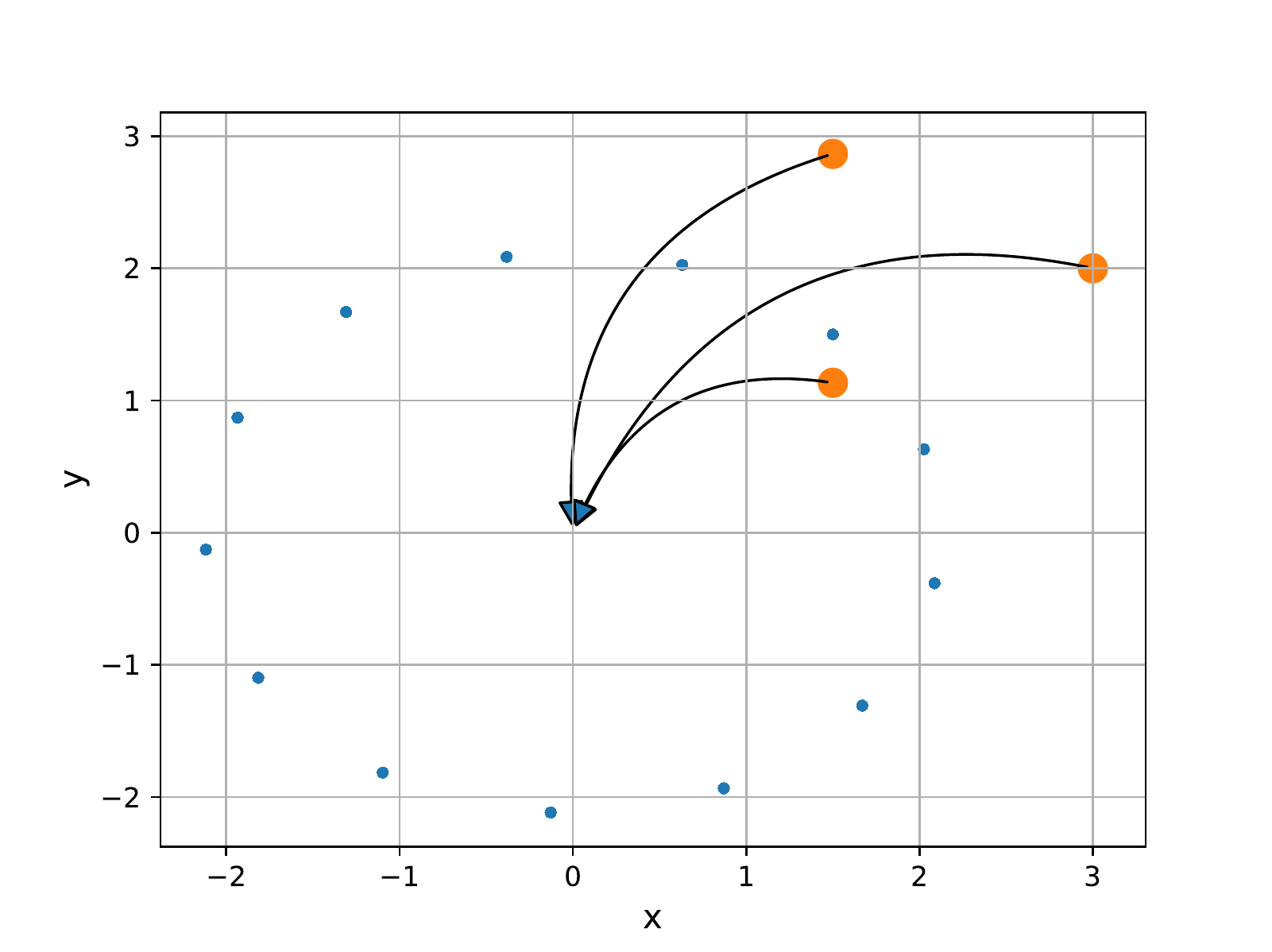}
	\includegraphics[width=0.495\linewidth]{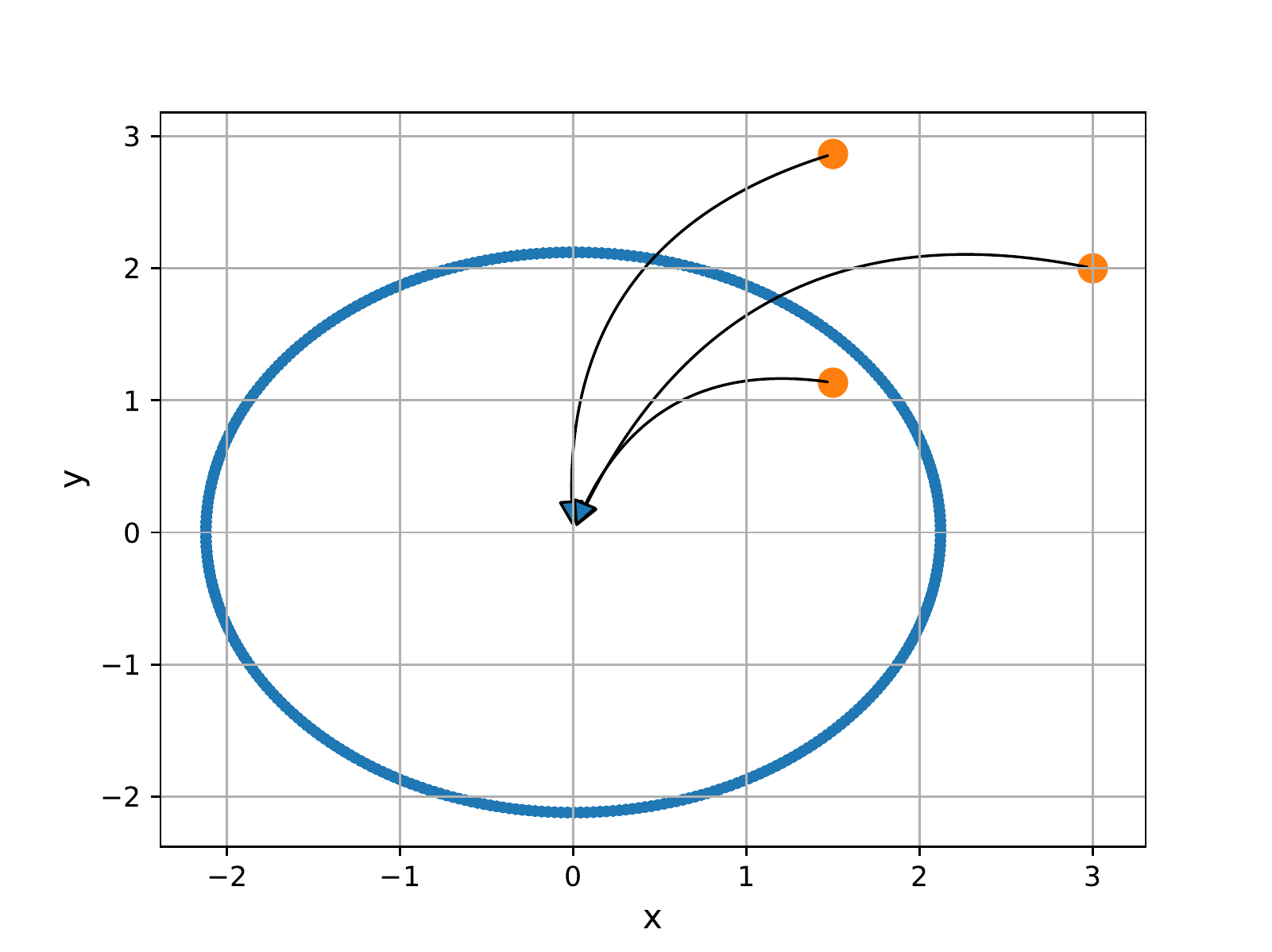}\\
  \caption{\label{Quantum_shapes_3} Simulation of quantum chaos game for bosonic particles. In both cases $\tau=1$ and $\lambda=3\pi$. The coherent states are indicated by the three larger circles of coordinates $\beta_j=e^{i\theta_j}+2(1+i)$ with $\theta_j=2\pi j/3$ $j=0,1,2$. Black arrow takes us from every $\beta_{j}$ to its transformations $\tilde\beta_{n,j}=0$. On the left $\omega=7\pi/13$, on the right $\omega=1$.  
  }
\end{figure}

For simplicity one can write $\omega\tau=\Omega$ and $\lambda\tau=\Lambda$. The analogy with the chaos game now becomes obvious (though as yet imperfect): $\chi_{n+1}=(1-W)\chi_{n}+W\tilde\beta_{n,\gamma_{n+1}}$, with
\Beq
W&=&1-\cos(\Lambda)e^{-i\Omega}\\
\tilde \beta_{n,\gamma_{n+1}}
&=&
\frac{ie^{-i (n+1)\Omega}\sin(\Lambda)}{1-\cos(\Lambda)e^{-i \Omega}}\beta_{\gamma_{n+1}}.
\Eeq
If $\Omega$ is a multiple of $2\pi$, one perfectly recover the $2d$ chaos game as presented in section \ref{classical} with the position of the vertex being defined by $\tilde\beta_j=i\sin(\Lambda)\beta_j/(1-\cos(\Lambda))$. For example choosing $\Lambda=\pi/3$ and $\beta_j=e^{i\theta_j}$ with $\theta_j=2\pi j/3$ (for $j=0,1,2$) leads to $W=1/2$ and the Sierpinski triangle, with the position of the vertex of the triangle (in the complex plane) given by $\tilde\beta_j=i\sqrt{3}\beta_j$ (see figure \ref{Quantum_shapes_1}). There are several other cases of potential interest, which we plan to investigate more closely in subsequent work. 
\begin{enumerate}
\item If $\Omega=\pi$,  there are two distinct points $\tilde\beta_{n,j}$ for every $\beta_{j}$ (assuming $\beta_{j}\ne0$) such that $\tilde\beta_{2n,j}=-\tilde\beta_{2n+1,j}$. Effectively the chaos game is played between $2M$ vertices (see figure \ref{Quantum_shapes_1}).
\item In general if $\Omega=\pi r/s$ (with $r,s\in\Nbb$ and ${\rm gcd}(r,s)=1$) the quantum game is effectively a classical chaos game with $sM$ or $2sM$ vertices if $r$ is respectively even or odd (see figure \ref{Quantum_shapes_2}). 
\item If $\Omega/\pi$ is irrational, the pattern we observe is the result of a continuous rotation of the original fractal (see figure \ref{Quantum_shapes_2}).
\item If $\Lambda$ is a multiple of $\pi$ (say $\Lambda=\pi p$) then $\tilde \beta_{n,j}=0$ $\forall n\in\Nbb$. It follows that regardless of the state of the subsystem the system interacts with, one has 
$\chi_{n}=(-1)^{pn}e^{-i\Omega n}\chi_{0}$. The $\chi_{n}$ values are orbiting on the circle of radius $|\chi_{0}|$. The orbit is either discrete or continuous. If one consider $\Omega/\pi$ to be rational (say $\Omega=\pi r/s$ with $r,s\in\Nbb$ and ${\rm gcd}(r,s)=1$) then $\omega/\lambda$ is rational as well. It follows that the set $\{\chi_n\}$ is finite, with cardinality $|\{\chi_n\}|=s$ or $|\{\chi_n\}|=2s$ if $ps+r$ is respectively even or odd. However if $\Omega/\pi$ is irrational so is $w/\lambda$ and $\chi_{n}$ fall on a continuous orbit of radius $|\chi_{0}|$ (see figure \ref{Quantum_shapes_3}).
\end{enumerate}

\subsection{The absence of Entanglement}
Our work has now lead us to the following dilemma. On one hand, we are working with coherent states which are often refered to as states with the most classical-like behaviour. On the other hand, just as in the fermionic study, we would like to investigate the evolution of entanglement which is said to be a purely quantum effect. This lead us to the following question: Can entanglement emerge from the quantum evolution of classical-like states? In the following we argue that under the conditions considered here (quadratic Hamiltonians and initially factorised coherent state), the quantum dynamics does not allow for entanglement between the system and the bath.\\

Let us remind the reader of the equation $b_0\ket{n\tau}=\chi_n\ket{n\tau}$, reflecting the fact that the system is at any time in a coherent state. In fact, one can be more specific and show that at all time $t=n\tau$ the entire state can be written as a product of coherent states. Let us rewrite the state $\ket{\tau}$ generated after one iteration:
\Bee
\ket{\tau} = U_{(1)}(\tau)\ket{\chi_0}\bigotimes_{j\ge1}\ket{\gamma_j}_j=\ket{01(\tau)}\bigotimes_{j\ge2}\ket{\gamma_j(\tau)}_j.
\Eee
with $\ket{01(\tau)}=e^{-i\tau (H_0+V_{0,1}+H_1)}\ket{\chi_0}\otimes\ket{\gamma_1}$ and  $\ket{\gamma_j(\tau)}_j=e^{-i\tau H_j}\ket{\gamma_j}_j$.
It can be easily shown that $\ket{\gamma_j(\tau)}_j$ is the coherent state parametrised by complex number $e^{-i\omega\tau}\gamma_j$. Along the same lines, it is not hard to show that $\ket{01(\tau)}$ is itself the product of two coherent states. We write $\ket{01(\tau)}=\ket{\chi_1}_1\otimes\ket{\tilde \gamma_1(\tau)}$, where $\chi_1$ is given by equation \eqref{eq_chi_evol}. Reasoning by induction, it follows that at all time $t=n\tau$ the ensemble system-environment remain in a product of coherent states. Even if the system and the bath influence each other, the states are not mixing, and only the parameters associated to each state is evolving. It follows that the system is and remains in a pure state at all times, leading to no entanglement.

\section{Conclusion}
In this paper, we have shown that repeated interactions of an appropriate quantum system with a bath provide a physical model of Barnsley's chaos game. We follow the evolution of the system after successive interactions with the subsystems of the environment. All subsystems are independent and randomly prepared in one of the predetermined $\ket{\beta_j}$ states.
When considering fermionic particles, we are limiting the set of possible $\ket{\beta_j}$s to the vacuum state $\ket{0}$ and the one particle state $\ket{1}$. We verify that the system's reduced density matrix is at all time in a Gaussian state. The effect of the repeated interactions onto the system is tracked by following the evolution of the expected number of particles $\NN_n$. This quantity is governed by the chaos game equation. It follows that for $\ww =\sin(\lambda\tau)>1/2$ the expected number of fermions fall onto the Cantor-like set. However, for $\ww<1/2$, $\NN_n$ takes continuous value in the interval $[0,1]$. In this case, we investigate the density of points in the limit $t\rightarrow\infty$. {Up to a translation and a dilatation, the distribution of point is identical to the Bernoulli convolutions presented in \cite{Krapisvsky_03, Turban_10}, in the context a the random walk with variable size steps.} {We propose an ansatz for small $x$, which allows us to find the $\ww$ values for which a $x^m$ behaviour is expected. Those $\ww$ appear to be given by the values for which the distribution is known to be absolutely continuous.} We track the evolution of entanglement between the system and the environment. Results from numerical simulations are compared to approximation using moment expansion. When considering bosonic modes, the $\ket{\beta_j}$ states are coherent states with parameters $\beta_j\in\Cbb$. The effect of the bath on the system can be directly observed by following the evolution of the eigenvalues $\chi_n$ of the annihilation operator $b_0$ at time $t=n\tau$. The eigenvalue $\chi_n$ is governed by a chaos game-like equation. In fact for specific parameter values (such that $\Omega$ is multiple of $2\pi$) the analogy with the chaos game is perfect. In the case of the Sierpinski triangle, the vertices of the triangle are not given by the $\beta_j$ values themselves, but by the transformations $\tilde\beta_j$. Other interesting scenarios emerge for different parameter subspaces. Roughly speaking, one generally observe either discrete or continuous rotation of a classical chaos game pattern. Finally, the bosonic set up is such that no entanglement emerges. We argued that the system remained at all time, in a direct product of coherent states.

\section*{Acknowledgements}
The authors would like to thanks the AMRC in Coventry and the Stat. Phys. Group in Nancy for their support and constant efforts to protect curiosity driven research. In particular, T. Platini extends his acknowledgements to S. Vantieghem, N. Fytas, D. Karevski and L. Turban.

\clearpage

\clearpage


\clearpage

\section{Appendix}

\subsection{Evolution of Fermionic Operator}
\label{fermionic_evolution}
This section presents the canonical method often used when solving quadratic fermionic systems. Here the time evolution is considered over one time step of the repeated interaction only. To prove equations \eqref{eq_evol_f_0} and \eqref{eq_evol_f_n} one can restrict ourselves to the subspace $\HH_0\otimes\HH_n$. Let us start by writing  $H_{0,n}=\Fbf_n^\dag\Tbf\Fbf_n$ with $\Fbf_n=(f_0,f_n)^T$ \footnote{For bosonic systems, follow the same procedure, replacing fermionic operators by bosonic operators: $f_j\rightarrow b_j$ and $\Fbf\rightarrow\Bbf$.}. We write $\phi_1$, $\phi_2$ and $\epsilon_1$, $\epsilon_2$ the eigenvectors and eigenvalues of $\Tbf$. We also define the matrix $\Ubf$ such that $\Ubf^\dag\Ubf=\Ubf\Ubf^\dag=\id$ and so that $\Dbf=\Ubf^\dag\Tbf\Ubf$ is diagonal: $\Dbf_{i,i}=\epsilon_{i}$.  It follows that the elements of $\Ubf$ are given by the components of the eigenvectors $\phi_i$: $\Ubf_{j,i}=\phi_i(j)$. We then define $\bar\Fbf=\Ubf^\dag\Fbf_n$ and $\bar\Fbf^\dag=\Fbf_n^\dag\Ubf$ and write $\bar\Fbf=(\bar f_1, \bar f_2)^T$. The fermionic anticommutation relations \footnote{Obviously, bosonic commutation relations have to be satisfied for the bosonic scenario.} are easily verified:
\Beq
\{ \bar f_j , \bar f_i^\dag\}=\delta_{i,j}, \  \{ \bar f_j , \bar f_i\}=\{ \bar f_j^\dag , \bar f_i^\dag\}=0.
\Eeq
Finally, we note that the Hamiltonian takes the so called free fermionic form: $H_{0,n}=\epsilon_1\bar f_1^\dag \bar f_1 +\epsilon_2\bar f_2^\dag \bar f_2$. The evolution of $\bar f_1$ and $\bar f_2$ is generated by the action of $\exp(-i\tau H_{0,n})=V_1(\tau)V_2(\tau)$ with $V_j(t)=\exp(-i\tau\epsilon_j\bar f_j^\dag \bar f_j)$. It follows that the evolution of $\bar f_j(\tau)$ is simply $\bar f_j(\tau)=\exp(-i\tau\epsilon_j)\bar f_j$. Equivalently we write $\bar \Fbf(\tau)=e^{- i\tau\Dbf}\bar\Fbf$, from which we derive $\Fbf_n(\tau)=e^{- i\tau\Tbf}\Fbf_n$, leading to equations \eqref{eq_evol_f_0} and \eqref{eq_evol_f_n}.

\subsection{Showing $\bra{n\tau}f_0\ket{n\tau}=0$}
\label{fermionic_null}
In this section we show that that $\bra{n\tau}f_0\ket{n\tau}=0$. Let us start with equation \eqref{eq20} which easily leads to
\Beq
\bra{n\tau}f_0\ket{n\tau}&=&\left(e^{-i\tau \Tbf}\right)_{1,1}\bra{(n-1)\tau}f_0\ket{(n-1)\tau}\nonumber\\
&+&\left(e^{-i\tau \Tbf}\right)_{1,2}\bra{(n-1)\tau}f_n\ket{(n-1)\tau}.
\Eeq
\begin{enumerate}
\item First  we evaluate $\bra{(n-1)\tau}f_n\ket{(n-1)\tau}$. We remind the reader that up to step $n-1$, the $n^{\rm th}$ subsystem of the environment has evolved freely under the iterated action of the operator $\exp(-i\tau H_n)$. Hence we can write
\Beq
\bra{(n-1)\tau}f_n\ket{(n-1)\tau}
&=& \hphantom{.}_n\bra{\gamma_n}e^{i(n-1)\tau H_n} f_n e^{-i(n-1)\tau H_n} \ket{\gamma_n}_n\nonumber \\
&=&\hphantom{.}_n\bra{\gamma_n} 
f_n((n-1)\tau)\ket{\gamma_n}_n.
\Eeq
Since $f_n(t)=e^{-it\omega}f_n$ (for all $t\le(n-1)\tau$) one has $\bra{(n-1)\tau}f_n\ket{(n-1)\tau}=e^{-i(n-1)\tau\omega}\hphantom{.}_n\bra{\gamma_n} f_n \ket{\gamma_n}_n$. The state $\ket{\gamma_n}_n$ being either $\ket{0}_n$ or $\ket{1}_n$, one immediately sees that $_n\bra{\gamma_n} f_n \ket{\gamma_n}_n=0$.
\item We are now left with
\Beq
\bra{n\tau}f_0\ket{n\tau}=\left(e^{-i\tau \Tbf}\right)_{1,1}\bra{(n-1)\tau}f_0\ket{(n-1)\tau},
\Eeq
which iteratively takes us to $\bra{n\tau}f_0\ket{n\tau}=[\left(e^{-i\tau \Tbf}\right)_{1,1}]^n\bra{I}f_0\ket{I}$ for which we have $\bra{I}f_0\ket{I}=\hphantom{.}_0\bra{0}f_0\ket{0}_0=0$. 
\end{enumerate}

\subsection{Evolution of $\NN_n$}
\label{fermionic_N_evolution}
Starting from equation \eqref{eq20} we have
\Beq
\NN_n&=&|(e^{-i\tau \Tbf})_{1,1}|^2\NN_{n-1}+|(e^{-i\tau \Tbf})_{1,2}|^2
\bra{(n-1)\tau}f_n^\dag f_n\ket{(n-1)\tau}\nonumber\\
&+&
(e^{-i\tau \Tbf})_{1,1}^*(e^{-i\tau \Tbf})_{1,2}
\bra{(n-1)\tau}f_0^\dag f_n\ket{(n-1)\tau}\nonumber\\
&+&
(e^{-i\tau \Tbf})_{1,2}^*(e^{-i\tau \Tbf})_{1,1}
\bra{(n-1)\tau}f_n^\dag f_0\ket{(n-1)\tau}.
\Eeq
Once again it is easy to verify that $\bra{(n-1)\tau}f_n^\dag f_n\ket{(n-1)\tau}=\hphantom{.}_n\bra{\gamma_n}f_n^\dag f_n\ket{\gamma_n}=\gamma_n$, while the cross term can be shown to vanish. In fact, up to time $(n-1)\tau$ the system and subsystem $n$ are independent. It follows that
\Beq
\bra{(n-1)\tau}f_0^\dag f_n\ket{(n-1)\tau}\propto
\hphantom{.}_n\bra{\gamma_n} 
f_n((n-1)\tau)\ket{\gamma_n}_n=0.
\Eeq
We are therefore left with equation \eqref{eqNNn}.

\end{document}